\documentclass[a4paper,11pt]{article}
\usepackage{jheppub} 
\makeatletter
\gdef\@fpheader{Prepared for submission to JHEP}
\usepackage{lineno}
\usepackage{bm}
\usepackage{graphicx}
\usepackage{subcaption}

\title{\boldmath Constraining electromagnetic couplings of ultralight scalars from compact stars}

\author[a,b]{Tanmay Kumar Poddar,}
\author[c]{and Amol Dighe}

\affiliation[a]{INFN, Gruppo collegato di Salerno,\\ Via Giovanni Paolo II 132, Fisciano I-84084, Italy}
\affiliation[b]{Institute for Particle Physics Phenomenology (IPPP), Department of Physics, Durham University, \\ Durham DH1 3LE, United Kingdom}
\affiliation[c]{Tata Institute of Fundamental Research,\\
Homi Bhabha Road, Colaba, Mumbai 400005, India}

\emailAdd{tanmay.k.poddar@durham.ac.uk}
\emailAdd{amol@theory.tifr.res.in}

\abstract{If an ultralight scalar interacts with the electromagnetic fields of a compact rotating star, then a long range scalar field is developed outside the star. The exterior scalar field profile is Coulomb-like in the far field region, corresponding to a scalar-induced charge associated with the star.
In a binary star system, the scalar-induced charge would result in a long range force between the stars, with the scalar field acting as the mediator.
The scalar-photon interactions would modify Maxwell's equations for electromagnetic fields in vacuum, however the scalar interaction would not change the photon dispersion relation at the leading and the next-to-leading geometric optics order. 
The scalar field would also induce additional electric and magnetic fields and hence affect the measurement of the magnetic field of the star at different latitudes. A scalar field sourced by  time-varying electromagnetic fields can also carry away energy from a compact star in the form of radiation,
and hence contribute to its spin-down luminosity.   
We constrain the scalar-photon coupling 
from the measurements of the spin-down luminosity
using the Crab pulsar, the soft gamma repeater SGR 1806-20, and the gamma ray burst GRB 080905A. 
We also project sensitivities on the coupling from future measurements of
the long-range force between two compact stars in a binary such as PSR J0737-3039 and binary magnetar systems, and from the estimates of magnetic field at the stellar surface.
Future advances in telescopes and targeted observations of stars with strong surface magnetic fields, large radii, and high spin frequency can substantially tighten these coupling limits.}

\begin{document}

\maketitle
\flushbottom

\section{Introduction}

Neutron stars (NSs) --- or pulsars --- act as remarkable cosmic laboratories for exploring the mysteries of the Universe. They play a crucial role in generating gravitational waves (GWs), as evidenced by the GW170817 event \cite{LIGOScientific:2017vwq} that has paved the way for advancements in multi-messenger astronomy \cite{Cowperthwaite:2017dyu}. These dense, rotating, magnetized objects emit radio waves so regularly that they behave like cosmic clocks. The typical mass of a NS is $1.4~M_\odot$ and its radius is $10-20~\mathrm{km}$. The magnetic field of the NS is dipolar and its strength is about $10^{12}~\mathrm{G}$ \cite{Gold:1968zf,Goldreich:1969sb,Sturrock:1971zc,Ruderman:1975ju,Weekes:1989tc,Lattimer:2004pg}. If the magnetic field is even stronger $(\gtrsim 10^{14}~\mathrm{G})$, then the compact object is called a magnetar \cite{Duncan:1992hi,Usov:1992zd,Stella:2005yz,Turolla:2015mwa,Kaspi:2017fwg,Beniamini:2017ilu,Esposito:2018gvp,CHIMEFRB:2020abu,Lin:2020vlo,DallOsso:2021xbv}. 

Compact stars (NSs and magnetars) also serve as probes to search for the dark matter (DM) in the Universe \cite{Raffelt:1999tx,Raffelt:2006cw,Bramante:2023djs,Caputo:2024oqc}. Results from the Planck satellite suggest that the energy density of DM is about five times that of the visible matter \cite{Planck:2018nkj}. The weakly-interacting massive particle (WIMP) motivated by the theory of supersymmetry has been one of the leading candidates for DM \cite{Jungman:1995df}. However, constraints on WIMPs from direct detection experiments \cite{Roszkowski:2017nbc,XENON:2018voc,XENON:2020gfr,LZ:2022lsv} and the small scale structure of the Universe \cite{Bullock:2017xww} motivate us to study alternative candidates for DM. 
Ultralight DM is one such promising candidate, where sub-eV mass range particles can account for the present DM relic density of the Universe, at the same time staying consistent with the direct search experiments and cosmological observations  \cite{Hu:2000ke,Marsh:2015wka,Hui:2016ltb,Robles:2018fur,Amin:2022pzv}. 
If such a DM candidate has mass as low as $10^{-22}~\mathrm{eV}$, its de Broglie wavelength would be of the order of the size of a dwarf galaxy $(1-2~\mathrm{kpc})$. The number density of ultralight DM within this de Broglie wavelength is $10^{30}/\mathrm{cm^3}$ for the local DM density $\rho_\odot\sim 0.4~\mathrm{GeV}/\mathrm{cm^3}$. The presence of such a large number density implies that DM oscillates coherently in a wave-like manner or exhibits long range behavior, potentially forming a Bose-Einstein condensate \cite{Boehmer:2007um,Li:2013nal,KumarPoddar:2019jxe,DiGiovanni:2022mkn,Davoudiasl:2024grq,Poddar:2024thb,Dutta:2024vzw}. The ultralight DM can be scalar \cite{Khmelnitsky:2013lxt,Hui:2016ltb,Badurina:2021lwr,Tretiak:2022ndx,Matos:2023usa}, pseudoscalar \cite{Zhang:2016uiy,Hook:2017psm,Hook:2018iia,Dror:2020zru,Chadha-Day:2021szb,Winch:2024mrt,Li:2024bbe}, vector \cite{Dror:2018pdh,PPTA:2022eul,Fedderke:2022ptm,Chase:2023puj,Chowdhury:2023xvy}, or tensor \cite{Brito:2020lup,Wu:2023dnp,Guo:2023gfc}; some such particles are also motivated from string/M theory \cite{Witten:1984dg,Battye:1994au,Yamaguchi:1998gx,Svrcek:2006yi,Arvanitaki:2009fg}. 

In addition to its gravitational interactions, if the DM interacts with the Standard Model (SM) particles with very small interaction strengths (allowed by the current data), then precision measurements at the existing and forthcoming experiments can either detect or constrain its properties. No observations or experiments have found the nature of DM so far. However, there are several tests which put constraints on ultralight DM,
for example, gravity tests \cite{Wagner:2012ui,KumarPoddar:2019jxe,KumarPoddar:2019ceq,KumarPoddar:2020kdz,Poddar:2020qft,Poddar:2021sbc,Poddar:2021ose,Poddar:2023pfj}, magnetometer searches \cite{Budker:2013hfa,Kim:2021eye}, Lyman-$\alpha$ observations \cite{Nori:2018pka,Rogers:2020ltq}, search for black hole (BH) superradiance \cite{Arvanitaki:2010sy,Baryakhtar:2017ngi,Brito:2017zvb}, variation of fundamental constants \cite{Flambaum:2007my,Stadnik:2014tta,Leefer:2016xfu,Kaplan:2022lmz,Oswald:2021vtc,Brzeminski:2022sde,Sherrill:2023zah,Bloch:2023uis}, cosmic microwave background (CMB) observations \cite{Hlozek:2014lca,Hlozek:2017zzf}, and more \cite{Joshipura:2003jh,Bandyopadhyay:2006uh,Huang:2018cwo,Dror:2020fbh,Alonso-Alvarez:2021pgy,Gherghetta:2023myo,Alonso-Alvarez:2023tii,Arvanitaki:2014wva,Arvanitaki:2016qwi,Kopp:2018jom,Tsai:2021irw,Dentler:2021zij,Dalal:2022rmp}. 
The existing bounds on the coupling of ultralight DM with photons, as determined from different experiments, are summarized in \cite{AxionLimits}. 

The phenomenology of ultralight scalar, pseudoscalar, vector, and tensor fields is remarkably diverse, and numerous studies over the years have explored their potential signatures in cosmic laboratories. The dilaton and axion field profiles in different string gravity models of BH have been discussed in \cite{Campbell:1990ai,Campbell:1991kz,Campbell:1991rz,Mignemi:1992pm}, where the field is sourced by the Chern-Simon and Gauss-Bonnet terms. In this paper, we consider the scenario where an ultralight scalar $\phi$ (which need not be the DM) interacts with the CP-even electromagnetic (EM) current $F_{\mu\nu}F^{\mu\nu}$. As we shall show later,
such an interaction, in the presence of the large EM fields near the surface of a rotating compact star, leads to a long range scalar field,\footnote{Note that ultralight pseudoscalars such as axions may interact with the charge-parity (CP)-odd EM current $F_{\mu\nu}\tilde{F}^{\mu\nu}$. However, the resultant pseudoscalar field goes as $a \sim \cos\theta/r^2$ \cite{Mohanty:1993nh,Chadha-Day:2026ban}, where $\theta$ is the polar angle for a rotating magnetized compact star, i.e., it falls faster than the scalar field as one moves away from the source. The influence of an axion background on EM radiation is examined in \cite{Harari:1992ea,Krasnikov:1996bm,McDonald:2019wou,Sokolov:2022fvs,Domcke:2023bat,Chadha-Day:2026ban}.
}
$\phi \sim 1/r$ in the far field region. This scalar field, in turn, induces additional electric and magnetic fields around the source. We explore four kinds of effects of such a scalar field:

\begin{itemize}
\item  The scalar field sourced by the EM field of a compact star can influence the propagation of photons through the resulting scalar background. To investigate this effect, we employ the eikonal approximation to study the propagation of photons emitted by pulsars in the presence of the scalar field.

\item In a binary system of two compact stars, an ultralight scalar particle can mediate a long-range force in addition to the gravitational force between the stars. Various fifth-force experiments can place constraints on such long-range interactions. 

\item The scalar field sourced by the EM field of the compact star modifies the star's background magnetic field through its interaction with the EM sector. As a result, the magnetic field outside the star acquires a scalar-induced correction, leading to deviations from the standard magnetic field profile at different altitudes.

\item  If the source is time-dependent, the scalar field itself can also act as a form of radiation, carrying away energy from the compact star. This leads to a decrease in their spin rate, a process known as spin-down. 

\end{itemize}
The measurements of observables corresponding to the above effects would allow us to constrain the scalar-photon coupling. 

In this paper, we do not explore the detailed mechanisms of scalar mass generation, since our focus is not on model building or ultraviolet completion.
However, it is known that such ultralight scalars can naturally acquire mass in several theoretical frameworks. These include clockwork mechanism \cite{Giudice:2016yja,Wood:2023lis}, Planck-suppressed operators from string theory or quantum gravity \cite{Hui:2016ltb,Hubisz:2024hyz,Banerjee:2022wzk}, and non-perturbative effects such as instantons \cite{Kitano:2021fdl}. Ultralight scalars also commonly appear in extra-dimensional compactifications \cite{Anchordoqui:2023tln}, or scale-invariant symmetry breaking scenarios \cite{Ferreira:2020fam}.

The paper is organized as follows. In Section~\ref{sec1}, we derive the scalar field profile outside the compact star arising from the scalar-photon interaction, with the scalar field being sourced by the star's EM field. The scalar-induced magnetic field is calculated in Section~\ref{sec2}. In Section~\ref{sec3}, we calculate the photon propagation in presence of the scalar-photon interaction whereas the energy loss due to the scalar radiation sourced by the time-dependent EM field is derived in Section~\ref{sec4}. In Section~\ref{sec5}, we obtain constraints on the scalar-photon coupling from the spin-down measurements and obtain sensitivities on the strength of the coupling based on the searches for new long range force in a double pulsar binary and search for a scalar-induced magnetic field. Finally, in Section~\ref{sec6} we conclude and discuss our results.

We use the system of units with the speed of light in vacuum $c =1$, the reduced Planck constant $\hbar=1$,
and the Newton's gravitational constant $G=1$ throughout the paper, unless stated otherwise.

\section{Scalar field profile outside a compact star}\label{sec1}

A rotating compact star like a NS or a magnetar is a large dipole magnet. In the aligned rotator model (where the magnetic dipole moment is along the rotation axis of the star), the external dipolar magnetic field is given by
\cite{Goldreich:1969sb,Shapiro:1983du,Mohanty:1993nh}
\begin{equation}
\mathbf{B}^\mathrm{out}_{(r>R)}=B_0 R^3\Big(\frac{\cos\theta}{r^3}\hat{r}+\frac{\sin\theta}{2r^3}\hat{\theta}\Big) \, ,
\label{eq:1m}
\end{equation}
 where $R$ denotes the radius of the star, $B_0$ denotes the magnetic field strength at the magnetic pole on the stellar surface, and $\theta$ denotes the polar angle, which is measured with respect to the rotation axis of the star. 
Using the boundary condition that the tangential component of the electric field is continuous at $r=R$ while the normal component of the electric field may be discontinuous across the boundary, the expression for the electric field profile outside the star is \cite{Goldreich:1969sb,Shapiro:1983du,Mohanty:1993nh}
\begin{equation}
\mathbf{E}^\mathrm{out}_{(r>R)}=-\frac{B_0\Omega R^5}{2r^4}\Big[(3\cos^2\theta-1)\hat{r}+\sin2\theta\, \hat{\theta}\Big] \, ,
\label{eq:2m}
\end{equation}
where $\Omega$ denotes the angular velocity of the star. We assume that the compact star carries no net electric charge, so that no monopolar Coulomb contribution to the exterior electric field is present. Using Eqs. \eqref{eq:1m} and \eqref{eq:2m}, we can estimate the quantity $\frac{1}{2}F_{\mu\nu}F^{\mu\nu}=\mathbf{B}^2-\mathbf{E}^2$ outside the star as 
\begin{equation}
\mathbf{B}^2-\mathbf{E}^2=\frac{B^2_0R^6}{4r^6}(3\cos^2\theta+1)-\frac{B^2_0\Omega^2R^{10}}{4r^8}(5\cos^4\theta-2\cos^2\theta+1) \, ,
\label{eq:3m}
\end{equation}
where $F^{\mu\nu}$ denotes the EM field strength tensor.

In the following, we calculate the scalar profile outside a compact star, sourced by the exterior EM field while neglecting possible contributions from the stellar interior. To study the scalar field profile sourced by the EM fields outside a compact star, we write the Lagrangian for a CP-even scalar field interacting with the EM field as \cite{Domcke:2023bat}
\begin{equation}
\mathcal{L}=\frac{1}{2}\partial_\mu\phi\partial^\mu\phi-\frac{1}{4}F_{\mu\nu}F^{\mu\nu}-\frac{1}{4}g_{\phi\gamma\gamma}\phi F_{\mu\nu}F^{\mu\nu},
\label{eq:4m}
\end{equation}
where $\phi$ denotes the scalar field and $g_{\phi\gamma\gamma}$ denotes the effective coupling of this scalar field with the EM fields of the star. 
Note that $g_{\phi\gamma\gamma}$ can have either sign, as determined by the nature of the charged particles running in the triangle loop diagrams responsible for the effective scalar-photon interaction. When the loop contribution is dominated by fermions such as leptons or heavy quarks, the coupling $g_{\phi\gamma\gamma}$ is positive. However, if the loop is dominated by a $W$ boson, the coupling would acquire a negative sign \cite{Flambaum:2024zyt}.

Thus, the equation of motion of the scalar field can be obtained as  
\begin{equation}
\Box\phi=-\frac{1}{2}g_{\phi\gamma\gamma}(\mathbf{B}^2-\mathbf{E}^2) \, ,
\label{eq:5m}
\end{equation}
where the d'Alembertian operator in the Minkowski spacetime is $\Box=\frac{\partial^2}{\partial t^2}-\nabla^2$. Therefore, to have a non-trivial scalar field profile outside the compact star, we must have a nonzero ``source charge density''
$\rho_\phi = g_{\phi\gamma\gamma} (\mathbf{B}^2-\mathbf{E}^2)/2$ outside the star. 

We solve Eq.~\eqref{eq:5m} using the Green's function method and obtain the scalar field solution outside the star as (see Appendix \ref{app})
\begin{align}
\begin{split}
\phi(r,\theta)&=-g_{\phi\gamma\gamma}B^2_0\Bigg\{\Big[\frac{R^3}{12r}-\frac{R^6}{48r^4} -\frac{\Omega^2 R^5}{30 r}+\frac{\Omega^2 R^{10}}{180 r^6}\Big] +\\
&\Big[\frac{R^5}{20r^3}-\frac{R^6}{24r^4}-\frac{4\Omega^2 R^7}{315r^3}+\frac{\Omega^2R^{10}}{126 r^6}\Big] P_2(\cos\theta)
+\Big[-\frac{\Omega^2R^9}{63r^5}+\frac{\Omega^2 R^{10}}{70 r^6}\Big]P_4(\cos\theta)\Bigg\},
\end{split}
\label{total-main-text}
\end{align}
where the boundary conditions require that both the scalar field and its first radial derivative remain continuous across the NS surface, while the scalar field vanishes asymptotically at spatial infinity.

For a slowly rotating NS where the angular velocity is not very large, i.e. $|\mathbf{B}|\gg |\mathbf{E}|$, we can neglect the $\mathbf{E}^2$ term and
obtain the scalar field profile exterior to the star $(r>R)$ as 
\begin{equation}
\phi(r,\theta)\simeq -g_{\phi\gamma\gamma}B^2_0\Big[\frac{R^3}{12r}-\frac{R^6}{48r^4}+\Big(\frac{R^5}{20r^3}-\frac{R^6}{24r^4}\Big)P_2(\cos\theta)\Big].  
\label{tot-sc-field}
\end{equation}
The far field solution of the scalar profile to the leading order is
\begin{equation}
\phi(r,\theta)\simeq -\frac{g_{\phi\gamma\gamma}B^2_0R^3}{12r}+\frac{g_{\phi\gamma\gamma}B^2_0\Omega^2R^5}{30r}, 
\label{dur20}
\end{equation}
which for a slowly rotating star becomes 
\begin{equation}
\phi(r)\simeq \frac{Q_\phi^{\mathrm{eff}}}{r}+\mathcal{O}\Big(\frac{1}{r^3}\Big) \, ,
\label{gk2}
\end{equation}
where the effective scalar charge $Q_\phi^{\mathrm{eff}}$ is
\begin{equation}
Q_\phi^{\mathrm{eff}} \simeq-\frac{g_{\phi\gamma\gamma}B^2_0R^3}{12} \;.
\label{eq:nm1}
\end{equation}
Note that the scalar-induced effective charge, $Q^\mathrm{eff}_\phi$, can be either positive or negative, depending on the sign of the coupling $g_{\phi\gamma\gamma}$. This form of $Q^\mathrm{eff}_\phi$ arises from the term $F_{\mu\nu}F^{\mu\nu} \propto |\mathbf{B}|^2$ in the scalar source charge density $\rho_\phi$, under the assumptions of slow rotation of the star, a magnetically dominated regime $|\mathbf{B}| \gg |\mathbf{E}|$, and in the far region.

Equations~\eqref{gk2} and~\eqref{eq:nm1} indicate that a slowly rotating compact star develops a Coulomb-like long range scalar field associated with an effective scalar charge $Q_\phi^{\rm eff}$. Although these results are derived in the massless scalar limit, they remain applicable for $m_\phi r\ll 1$. For a finite scalar mass, the Coulomb-like contribution becomes suppressed at distances $r\gtrsim m_\phi^{-1}$. However, since the EM source extends throughout the exterior region, the complete massive scalar profile need not exhibit a purely Yukawa-suppressed behavior and requires solving the corresponding massive field equation. In particular, the near-surface massless approximation is valid for $m_\phi R\ll 1$.

We use Crab pulsar as a benchmark and obtain the absolute value of the scalar-induced charge from Eq.~\eqref{eq:nm1} as
\begin{equation}
Q_\phi^{\mathrm{eff}}\simeq 3.3\times 10^{29}\Big(\frac{g_{\phi\gamma\gamma}}{ 10^{-16}~\mathrm{GeV^{-1}}}\Big)\Big(\frac{B_0}{1.7\times 10^{13}~\mathrm{G}}\Big)^2\Big(\frac{R}{14~\mathrm{km}}\Big)^3,
\end{equation}
in the far field region outside of the star. Note that, for stars with a large angular velocity, the electric field outside the star cannot be neglected, and we need to solve the scalar field profile sourced by $\mathbf{B}^2-\mathbf{E}^2$ instead of only $\mathbf{B}^2$. Nevertheless, the inclusion of the additional $\mathbf{E}^2$ term does not alter the leading $1/r$ scaling of the far field scalar profile, it only modifies the corresponding effective monopole charge. A detailed analysis of this scenario is presented in Appendix \ref{app}.

We obtain the scalar field profile at the stellar surface from Eq.~\eqref{total-main-text} as (see also Appendix \ref{app})
\begin{equation}
\phi(R,\theta)=-g_{\phi\gamma\gamma}B^2_0R^2\Big[\frac{1}{16}+\frac{1}{120}P_2(\cos\theta)-(\Omega R)^2\Big(\frac{1}{36}+\frac{1}{210}P_2(\cos\theta)+\frac{1}{630}P_4(\cos\theta)\Big)\Big].   
\label{surface4}
\end{equation}
which for a slowly rotating star becomes
\begin{equation}
\phi(R,\theta)=  -g_{\phi\gamma\gamma}B^2_0R^2\Big(\frac{7}{120}+\frac{1}{80}\cos^2\theta\Big) .
\label{scalarsurfacemain1}
\end{equation}
Equation~\eqref{scalarsurfacemain1} shows that the field value at the surface of the slowly rotating star is not spherically symmetric. Thus, the absolute value of the scalar field is enhanced along the rotation/magnetic axis (for the aligned rotator model), reaching its maximum at the poles $(\theta=0,\pi)$, while it is minimized at the magnetic equator $(\theta=\pi/2)$. The absolute value of the scalar field value at the pole is roughly $21.4\%$ larger than the equatorial value. For a Crab pulsar with surface magnetic field $B_0\simeq 1.7\times 10^{13}~\mathrm{G}$ and radius $R\simeq 14~\mathrm{km}$, the field values at the pole and at the equator are $|\phi_{\mathrm{pole}}|\simeq 4\times 10^{10}~\mathrm{GeV}$ and $|\phi_{\mathrm{eq}}|\simeq 3.3\times 10^{10}~\mathrm{GeV}$, respectively for $g_{\phi\gamma\gamma}\simeq 10^{-15}~\mathrm{GeV}^{-1}$.

\begin{figure}
\includegraphics[width=15cm]{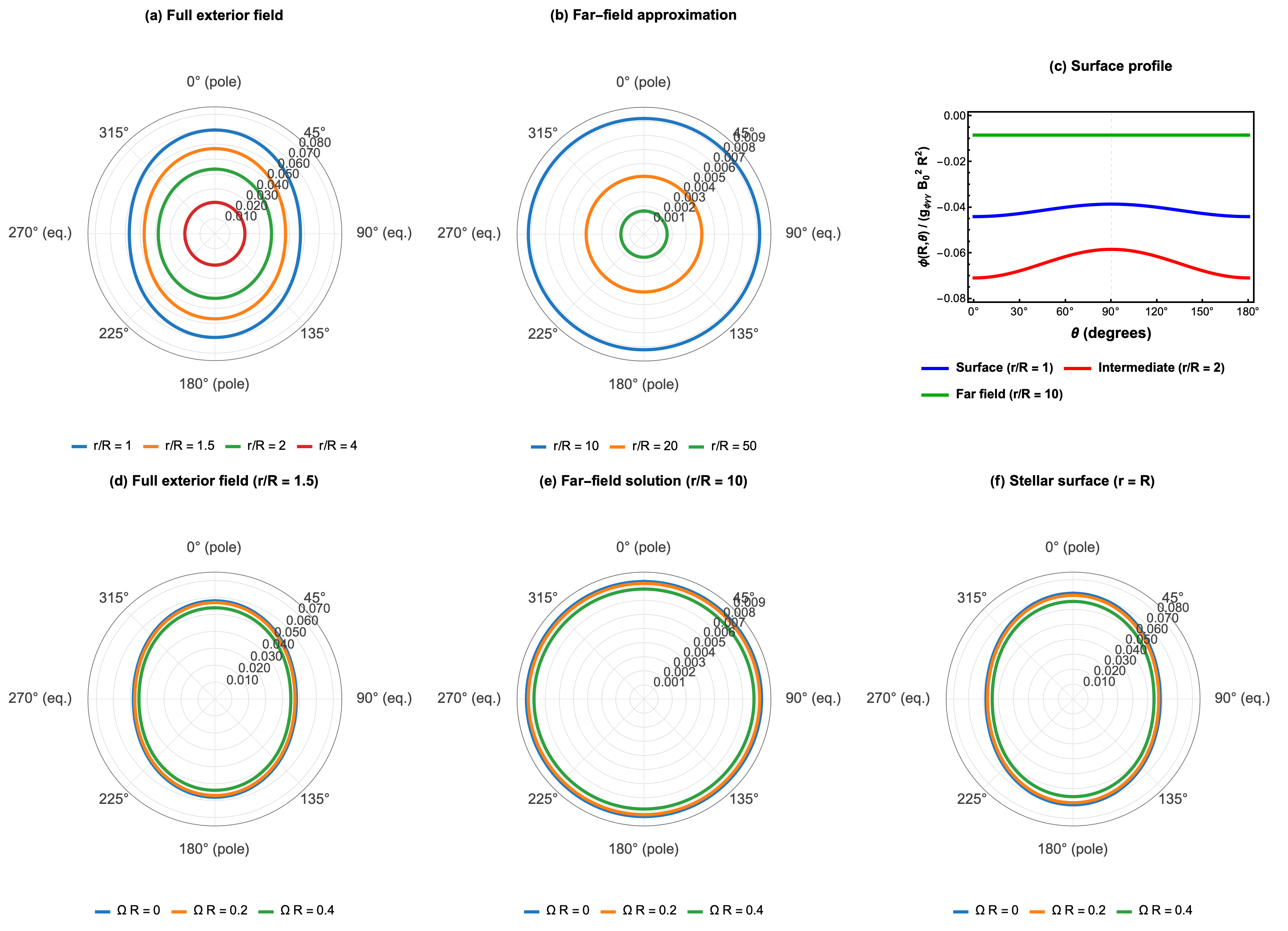}
\caption{
Panels (a) and (b) present the angular dependence of the magnitude of the scalar field $|\phi(r,\theta)|$, in units of $|g_{\phi\gamma\gamma} B_0^2R^2|$, as polar plots. Panel (c) shows the same quantity as a function of the latitude $\theta$. Panels (d), (e), and (f),
while plotting the same quantity on a polar plot, demonstrate the effect of rotation of the star.
}
\label{newfig}
\end{figure}

Figure~\ref{newfig} illustrates the radial and angular dependence of the absolute value of the scalar field generated by the NS, in units of $g_{\phi\gamma\gamma}B_0^2R^2$. Panel (a) shows the full exterior scalar field profile (cf. Eq.~\eqref{total-main-text}) at different radial distances, $r/R=1,\,1.5,\,2$, and $4$, with fixed $\Omega R=0.2$. The scalar field is strongest near the stellar surface and decreases with increasing distance from the NS. The angular distribution exhibits a clear polar enhancement, with the field attaining its maximum near the magnetic poles ($\theta=0^\circ,180^\circ$) and minimum near the equatorial plane ($\theta=90^\circ,270^\circ$). Increasing $r/R$ values take the field profile closer to spherical symmetry.

Panel (b) shows the far field approximation (cf. Eq.~\eqref{dur20}) at $r/R=10,\,20$, and $50$ for fixed $\Omega R=0.2$, where the absolute value of the scalar field amplitude decreases approximately as $1/r$, and the angular dependence becomes progressively weaker, approaching an almost spherically symmetric configuration.

Panel (c) explicitly shows the angular variation of the signed scalar field at the stellar surface (cf. Eq.~\eqref{scalarsurfacemain1}). The peak of the curve corresponds to the equatorial value of the field whereas the two extremes correspond to the field values at the poles. Overall, the results demonstrate that the scalar field exhibits a pronounced angular dependence near the NS surface, while its amplitude decreases with radial distance and its far-field profile becomes increasingly isotropic.

Finally, panels (d), (e), and (f) illustrate the effect of stellar rotation on the absolute value of the scalar field profile in the near field (cf. Eq.~\eqref{total-main-text}) region ($r/R=1.5$), in the far-field (cf. Eq.~\eqref{dur20}) region ($r/R=10$), and at the stellar surface (cf. Eq.~\eqref{surface4}) ($r=R$), respectively, for different values of the dimensionless rotational parameter $\Omega R=0,\,0.2$, and $0.4$. Increasing $\Omega R$ produces a modest suppression of the scalar field amplitude while largely preserving its angular distribution. The rotational dependence is therefore relatively weak compared with the overall radial attenuation of the field for the parameters considered.

\section{Scalar-induced EM fields from Maxwell's equations}\label{sec2}

The interaction of a CP-even scalar with the EM fields of the star modifies Maxwell's equations for the EM fields in vacuum. 
We derive the scalar-induced electric and magnetic field equations in a perturbative way by expanding the field strength tensor in powers of $g_{\phi\gamma\gamma}$, such that 
\begin{equation}
F^{\mu\nu}=F^{\mu\nu}_{(0)}+ F^{\mu\nu}_\phi;~~|F^{\mu\nu}_\phi|\ll |F^{\mu\nu}_{(0)}|,
\end{equation}
where ``$(0)$'' corresponds to any quantity in the limit $g_{\phi\gamma \gamma} =0$ and we retain terms to first order in $g_{\phi\gamma\gamma}\phi$ and in the induced EM field $ F^{\mu\nu}_\phi$. Since the scalar field itself is sourced by the EM background and satisfies $\phi=\mathcal{O}(g_{\phi\gamma\gamma})$, the induced EM correction is $\mathcal{O}(g^2_{\phi\gamma\gamma})$. The equation of motion of $F^{\mu\nu}_\phi$ in the covariant form is
$\partial_\mu F^{\mu\nu}_\phi=-g_{\phi\gamma\gamma}(\partial_\mu \phi)F^{\mu\nu}_{(0)}+\mathcal{O}(g^4_{\phi\gamma\gamma})$, which is in fact proportional to $g^2_{\phi\gamma\gamma}$, as $\phi\propto g_{\phi\gamma\gamma}$,
in the absence of source charge and current density of the plasma. This relation gives the expressions for the scalar induced electric $(\mathbf{E}_\phi)$ and magnetic $(\mathbf{B}_\phi)$ fields in terms of the background electric $(\mathbf{E}_{(0)})$ and magnetic $(\mathbf{B}_{(0)})$ fields as  
\begin{eqnarray}
\nabla\cdot \mathbf{E}_\phi & = & -g_{\phi\gamma\gamma} \, \mathbf{E}_{(0)} \cdot \nabla\phi \, , \nonumber \\
\nabla\times \mathbf{B}_\phi & = & \frac{\partial \mathbf{E}_\phi}{\partial t}-g_{\phi\gamma\gamma} \, \nabla\phi\times\mathbf{B}_{(0)} 
+ g_{\phi\gamma\gamma} \left(\frac{\partial \phi}{\partial t} \right) \mathbf{E}_{(0)}
\,,
\label{eq:19}
\end{eqnarray}
while the Bianchi identity $\partial_\mu \tilde{F}^{\mu\nu}_\phi=0$ gives
\begin{eqnarray}
\nabla\cdot \mathbf{B}_\phi & = & 0 \, , \nonumber \\
\nabla\times \mathbf{E}_\phi & = & -\frac{\partial \mathbf{B}_\phi}{\partial t} \, .
\label{eq:20}
\end{eqnarray}

In the aligned rotator configuration, the background fields and the scalar source are stationary so that the terms $\partial\mathbf{E}_\phi/\partial t$, $\partial\mathbf{B}_\phi/\partial t$ and
$\partial \phi/\partial t$ vanish. We restrict ourselves to the stationary particular solution and neglect free radiative homogeneous solutions. The scalar-induced magnetic and electric fields are produced due to the interaction of background magnetic (Eq.~\eqref{eq:1m}) and electric (Eq.~\eqref{eq:2m}) fields with the scalar. 
For static background EM fields, Eqs.~\eqref{eq:19} and \eqref{eq:20} represent how these background EM fields are altered due to the interaction with the scalar field. Thus, $\mathbf{E}_\phi$ and $\mathbf{B}_\phi$ are the leading scalar backreaction terms on the background EM fields.

Combining Eqs.~\eqref{eq:19} and \eqref{eq:20}, we obtain the wave equations for the scalar-induced magnetic field solution as
\begin{equation}
 \begin{split}
 \Box \mathbf{B}_\phi=-g_{\phi\gamma\gamma}\nabla\times (\nabla\phi\times \mathbf{B}_{(0)}).
\label{eq:21}     
 \end{split}   
\end{equation}
The generic form of the scalar-induced magnetic field exterior to the slowly rotating star $r>R$ is obtained as (see Appendix \ref{appnew2})
\begin{equation}
\begin{split}
\mathbf{B}_\phi(r,\theta)=g^2_{\phi\gamma\gamma}B^3_0R^6\Bigg\{\Big[\frac{1}{48r^4}-\frac{R^2}{160r^6}(3\cos^2\theta-1)\Big]\cos\theta\,\hat{\bm{r}}+\\ \Big[\frac{1}{48r^4}-\frac{R^2}{160r^6}(1+3\cos^2\theta)\Big]\sin\theta\,\hat{\bm{\theta}}\Bigg\}. 
\end{split}
\label{ds}
\end{equation}
The behavior of the scalar-induced magnetic field $\mathbf{B}_\phi$ in the far-field solution is
\begin{equation}
\mathbf{B}_\phi=\frac{g^2_{\phi\gamma\gamma}B^3_0R^6}{48r^4}(\cos\theta\,\hat{\bm{r}}+\sin\theta\,\hat{\bm{\theta}}),  
\label{newapp25-text}
\end{equation}
where the radial and polar components of the particular scalar-induced magnetic field scale as $r^{-4}$ in the far field region compared to the background dipolar field, which scales as $r^{-3}$.

The scalar-induced magnetic field may also contain source free homogeneous contributions satisfying $\nabla\times\mathbf B_\phi^{(h)}=0$ and $\nabla\cdot\mathbf B_\phi^{(h)}=0$ in the current free exterior. The $l=1$ homogeneous mode exhibits the same $r^{-3}$ dipolar dependence as the background magnetic field and can be absorbed into a redefinition of the stellar dipole moment. Higher homogeneous multipoles, such as the $l=3$ (octupole) contribution scaling as $r^{-5}$, may similarly be incorporated into the corresponding intrinsic stellar multipole moments. Within the centred-dipole approximation, imposing equatorial reflection symmetry excludes even-$l$ homogeneous modes. In the following, we restrict our analysis to the particular solution sourced by the scalar interaction and set the remaining homogeneous contributions to zero as a simplifying boundary prescription. A complete determination of these contributions would require matching the exterior solution to the EM configuration inside the star.

The scalar-induced electric field $\mathbf{E}_\phi$ can also be calculated similarly. Since the scalar-induced EM fields scale quadratically with $g_{\phi\gamma\gamma}$, and $\mathbf{B}_\phi$ is additionally suppressed by a factor $1/r$ compared to the background field at the far field limit, the deviations of the EM fields from their background values, 
as a result of their interactions with the scalar,
are small.

We calculate the scalar-induced magnetic field of a slowly rotating star at the stellar surface using Eq.~(\ref{ds}) (see also Appendix~\ref{appnew2}) as
\begin{equation}
\mathbf{B}_\phi(R,\theta)= \frac{g^2_{\phi\gamma\gamma}B^3_0R^2}{480}[(13-9\cos^2\theta)\cos\theta\,\hat{\bm{r}}+(7-9\cos^2\theta)\sin\theta\,\hat{\bm{\theta}}], 
\label{final-scalar-mag}
\end{equation}
which shows that the scalar-induced surface magnetic field is not spherically symmetric.

\begin{figure}
\includegraphics[width=15cm]{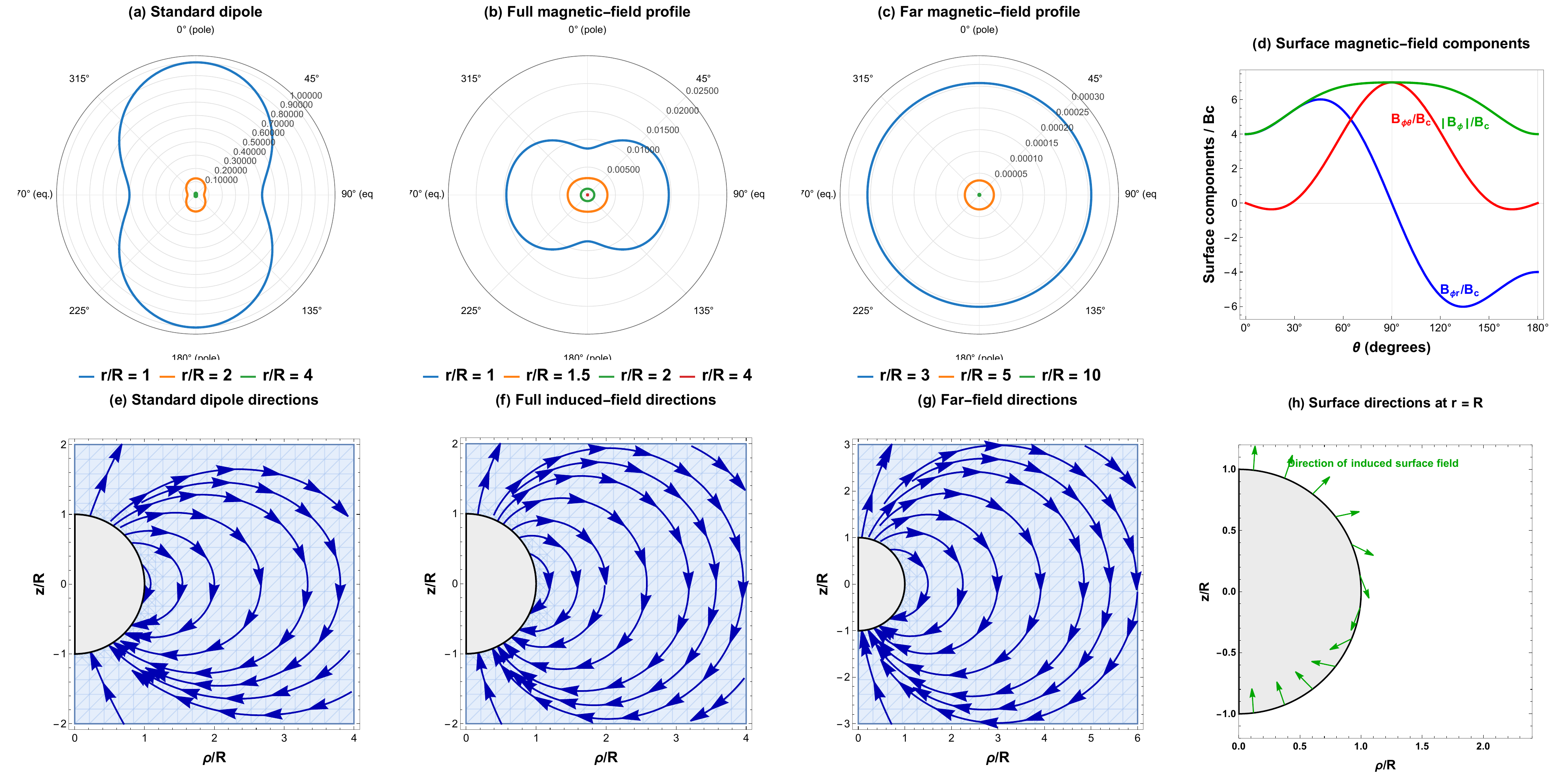}
\caption{Scalar-induced magnetic field profile around a compact star. Panels (a)-(c) show the magnitudes of the standard dipolar field, the full scalar-induced magnetic field, and its far field approximation, respectively, at different radial distances. Panel (d) shows the radial and polar components of the scalar-induced magnetic field and its magnitude at the stellar surface. Panel (e) denotes the standard dipole magnetic field direction and panels (f)-(g) illustrate the corresponding induced magnetic field directions in the meridional plane, while panel (h) shows the direction of the scalar-induced field along the stellar surface. Here, $\rho/R$ and $z/R$ denote the normalized cylindrical coordinates $(\rho=r\sin\theta, z=r\cos\theta)$, and the arrows indicate the local magnetic field direction. The magnetic field magnitudes are normalized by $B_0$ in panel (a), $g_{\phi\gamma\gamma}^2B_0^3R^2$ in panels (b)–(c), and $B_c=g_{\phi\gamma\gamma}^2B_0^3R^2/480$ in panel (d).}
\label{scalar-mag}
\end{figure}

Figure~\ref{scalar-mag} illustrates the spatial distribution and directional structure of the induced magnetic field in the exterior region of the NS, together with a comparison to the standard magnetic dipole configuration. Panel (a) shows the angular dependence of the standard dipole magnetic field strength (normalized with $B_0$) at different radial distances, $r/R=1,\,2$, and $4$ (cf. Eq.~\eqref{eq:1m}). The field exhibits the characteristic dipolar structure, attaining its maximum near the magnetic poles and minimum near the equatorial plane, while its magnitude decreases rapidly with increasing distance from the NS. 

Panel (b) presents the full induced magnetic field profile $|\textbf{B}_\phi|$ (in units of $g^2_{\phi\gamma\gamma}B^2_0 R^2$) at $r/R=1,\,1.5,\,2$, and $4$ (cf. Eq.~\eqref{ds}). Unlike the standard dipole configuration, the induced magnetic field exhibits a different angular distribution, with a relatively enhanced field strength toward the equatorial region. The field amplitude decreases with radial distance, as expected from the spatial attenuation of the underlying scalar-induced EM response.
Panel (c) shows the far field profile (cf. Eq.~\eqref{newapp25-text}) of the induced magnetic field $|\textbf{B}_\phi|$ (in units of $g^2_{\phi\gamma\gamma}B^2_0 R^2$) at $r/R=3,\,5$, and $10$, where the angular variation becomes considerably weaker and the field strength distribution approaches an approximately isotropic configuration.

Panel (d) illustrates the angular variations of the net scalar-induced magnetic field profile (green curve) at the stellar surface (cf. Eq.~\eqref{final-scalar-mag}) whereas the $\hat{\bm r}$ and $\hat{\bm \theta}$ components are shown with the blue and red curves respectively. Here, $B_c=g^2_{\phi\gamma\gamma}B^3_0R^2/480$ is the constant prefactor in Eq.~\eqref{final-scalar-mag}. Interestingly, although the background dipolar magnetic field is strongest at the magnetic poles, the scalar-induced correction attains its largest magnitude near the magnetic equator $(\theta=\pi/2)$. This behavior arises from the derivative structure of the scalar-induced source (See Eq.~(\ref{eq:21})) and the resulting admixture of dipolar $(l=1)$ and octupolar $(l=3)$ components, rather than from a simple rescaling of the background dipole field.

The directional structure of the magnetic field is illustrated in panels (e)-(h), where $\rho/R$ and $z/R$ denote the normalized cylindrical coordinates. Panel (e) shows the familiar dipole field line configuration of a standard magnetic dipole (cf. Eq.~\eqref{eq:1m}), with field lines extending from one magnetic hemisphere toward the other. Panel (f) illustrates the direction of the scalar-induced magnetic field (cf. Eq.~\eqref{ds}) in the meridional plane. Compared with the standard magnetic dipole, the induced field exhibits a modified spatial structure arising from the different angular dependence of its radial and polar components. Panel (g) shows that the induced field retains a coherent, large-scale directional structure in the far field region (cf. Eq.~\eqref{newapp25-text}), despite the substantial reduction in its magnitude. Finally, panel (h) illustrates the local orientation of the induced magnetic field along the stellar surface (cf. Eq.~\eqref{final-scalar-mag}), demonstrating how its direction varies with latitude. Overall, the figure demonstrates that the induced magnetic field differs from the conventional dipolar configuration in both its angular distribution and field line orientation. Although the induced field becomes progressively weaker with increasing radial distance, its nontrivial directional structure persists into the far field region.

\section{EM wave propagation in 
the background of a long range scalar field}\label{sec3}

Maxwell's equations of electrodynamics for the propagation of light are modified due to the interactions of the CP-even scalar field $\phi$ with the
EM radiation fields. Consider a situation where light, i.e., an EM wave, is emitted by the source in the presence of the background static EM and scalar fields.
In the absence of any other source plasma charge and current densities and in the weak background regime $|g_{\phi\gamma\gamma}\phi|\ll1$, the Maxwell's equations become \cite{Domcke:2023bat}
\begin{eqnarray}
\nabla\cdot \mathbf{E} & = & -g_{\phi\gamma\gamma}\mathbf{E}\cdot \nabla\phi \, , \nonumber \\
\nabla\times \mathbf{B} & = & \frac{\partial \mathbf{E}}{\partial t}-g_{\phi\gamma\gamma}\nabla\phi\times \mathbf{B} \, , \nonumber \\
\nabla\cdot \mathbf{B} & = & 0 \, , \nonumber \\
\nabla\times \mathbf{E} & = & -\frac{\partial \mathbf{B}}{\partial t} \, ,
\label{gt1}
\end{eqnarray}
where the $\mathbf{E}$ and $\mathbf{B}$ are the electric and magnetic fields of the propagating EM wave.
Using Eq.~\eqref{gt1}, we obtain the equation for the magnetic component of the EM wave as
\begin{equation}
\begin{split}
\Box\mathbf{B}=-g_{\phi\gamma\gamma}\nabla\times (\nabla\phi\times \mathbf{B})=g_{\phi\gamma\gamma}[(\nabla\phi\cdot \nabla)\mathbf{B}+(\nabla^2\phi)\mathbf{B}-(\mathbf{B}\cdot \nabla)\nabla\phi].
\label{gt2}
\end{split}
\end{equation}

To investigate the effects of the CP-even scalar interaction on photon propagation, we write the Lagrangian in Eq.~(\ref{eq:4m}) as
\begin{equation}
\mathcal{L}=-\frac{1}{4}Z(\phi)F_{\mu\nu}F^{\mu\nu},~~~Z(\phi)=1+g_{\phi\gamma\gamma}\phi,
\end{equation}
and the exact source-free Maxwell's equation becomes
\begin{equation}
\partial_\mu[Z(\phi)F^{\mu\nu}]=0.    
\label{cov}
\end{equation}
We write the eikonal ansatz for the vector potential as $A^\mu=a^\mu(x)e^{iS(x)/\epsilon}$ with $k_\mu=\partial_\mu S(x)$, where $S(x)$ is the eikonal phase and the eikonal expansion parameter $\epsilon\sim 1/kL_\phi\ll 1$. Here, $L_\phi=|\phi|/|\nabla\phi|$ denotes the characteristic variation scale of the scalar background. For the far field profile $\phi\propto 1/r$, one has $L\sim r$ and hence the geometric optics condition reduces to $kr\gg 1$. Therefore, the EM field tensor becomes
\begin{equation}
F^{\mu\nu}=e^{iS/\epsilon}\Big[\frac{i}{\epsilon}(k^\mu a^\nu-k^\nu a^\mu)+\partial^\mu a^\nu-\partial^\nu a^\mu\Big]. 
\label{emtensor}
\end{equation}
Substituting Eq.~\eqref{emtensor} into Eq.~\eqref{cov} and collecting terms proportional to $\mathcal{O}(\epsilon^{-2})$, we can write the leading order equation as
\begin{equation}
Z(\phi)[k_\mu k^\mu a^\nu-k^\nu(k_\mu a^\mu)]=0.  
\label{as}
\end{equation}
For a transverse polarization, we impose $k_\mu a^\mu=0$, and obtain the photon dispersion relation from Eq.~\eqref{as} as
\begin{equation}
k_\mu k^\mu=0,    
\end{equation}
since $Z(\phi)\neq 0$, and $a^\mu\neq 0$. Therefore, the CP-even scalar background does not modify the leading order photon dispersion relation and remains null.

To see if there is any change in the photon dispersion relation at the first sub-leading order $\mathcal{O}(\epsilon^{-1})$, we write Eq.~\eqref{cov} in terms of $A^\mu$ as 
\begin{equation}
\Box A^\nu -\partial^\nu(\partial_\mu A^\mu)+(\partial_\mu \ln Z )(\partial^\mu A^\nu-\partial^\nu A^\mu)=0. 
\label{expr}
\end{equation}
In the following, we write all the terms that appear in the right hand side of Eq.~\eqref{expr} in terms of the the eikonal expansion parameter. We obtain the first and second derivatives of $A_\mu$ from the eikonal equation as
\begin{equation}
\partial_\mu A^\nu=e^{i s/\epsilon}\Big[\partial_\mu a^\nu+\frac{i}{\epsilon}k_\mu a^\nu\Big],~~~\Box A^\nu=e^{iS/\epsilon}\Big[-\frac{1}{\epsilon^2}k_\mu k^\mu a^\nu+\frac{i}{\epsilon}(2k^\mu \partial_\mu a^\nu+a^\nu\partial_\mu k^\mu)+\Box a^\nu\Big].
\label{first-eikonal}
\end{equation}
We also obtain
\begin{equation}
\partial^\nu(\partial_\mu A^\mu)=e^{iS/\epsilon}\Big[-\frac{k^\nu(k_\mu a^\mu)}{\epsilon^2}+\frac{i}{\epsilon}(\partial^\nu (k_\mu a^\mu)+k^\nu \partial_\mu a^\mu)+\partial^\nu (\partial_\mu a^\mu)\Big].    
\label{second-eikonal}
\end{equation}
The scalar gradient contribution can be written as
\begin{equation}
(\partial_\mu \ln Z)(\partial^\mu A^\nu-\partial^\nu A^\mu)=e^{iS/\epsilon}\Big[\frac{i}{\epsilon}(\partial_\mu \ln Z)(k^\mu a^\nu-k^\nu a^\mu)+\mathcal{O}(\epsilon^0)\Big]. 
\label{third-eikonal}
\end{equation}
Thus, substituting Eqs.~\eqref{first-eikonal}, \eqref{second-eikonal}, and \eqref{third-eikonal} into Eq.~\eqref{expr} and collecting terms proportional to $\mathcal{O}(\epsilon^{-1})$, we get 
\begin{equation}
2k^\mu\partial_\mu a^\nu+a^\nu \partial_\mu k^\mu+a^\nu k^\mu \partial_\mu \ln Z-k^\nu(\partial_\mu a^\mu+a^\mu\partial_\mu \ln Z)=0.    
\label{polari1}
\end{equation}
Now we decompose the vector amplitude as $a^\mu=\mathcal{A}e^\mu$, where $\mathcal{A}$ is the slowly varying scalar amplitude and $e^\mu$ is the transverse polarization vector satisfying $k_\nu e^\nu=0$. Also, for a real normalized physical polarization, $e_\nu e^\nu=1$. Thus, Eq.~\eqref{polari1} becomes
\begin{equation}
2k^\mu\partial_\mu\mathcal{A}+\mathcal{A}\partial_\mu k^\mu+\mathcal{A}k^\mu\partial_\mu \ln Z=0,
\label{nwq1}
\end{equation}
which is the amplitude transport equation. In the limit $|g_{\phi\gamma\gamma}\phi|\ll 1$, Eq.~\eqref{nwq1} becomes
\begin{equation}
k^\mu\partial_\mu \ln \mathcal{A}=-\frac{1}{2}\partial_\mu k^\mu-\frac{1}{2}g_{\phi\gamma\gamma}k^\mu\partial_\mu \phi , 
\label{nwq2}
\end{equation}
which can be rewritten as
\begin{equation}
\partial_\mu[Z(\phi)\mathcal{A}^2k^\mu]=0. 
\label{nwq3}
\end{equation}
For a radial spherical wave propagation in the flat spacetime, we have $k^\mu=(\omega,k,0,0)$ and the conservation equation becomes
\begin{equation}
\frac{1}{r^2} \frac{d}{dr}[r^2 Z(r)\mathcal{A}^2(r)k]=0.   
\label{nwq4}
\end{equation}
At leading geometric optics order $k\simeq \omega=\mathrm{constant}$ and hence Eq.~\eqref{nwq4} results $r^2Z(r)\mathcal{A}^2(r)=\mathrm{constant}$, which gives
\begin{equation}
\mathcal{A} (r)\propto \frac{1}{r\sqrt{1+g_{\phi\gamma\gamma}\phi(r)}}\propto \frac{1}{r}e^{-g_{\phi\gamma\gamma}\phi(r)/2}+\mathcal{O}[(g_{\phi\gamma\gamma}\phi)^2],
\label{nwq5}
\end{equation}
for $|g_{\phi\gamma\gamma}\phi|\ll 1$. The $1/r$ factor is just the ordinary geometric dilution of a spherical wave; the extra $Z(\phi)^{-1/2}$ factor is the scalar-induced amplitude modification. Again, at this order $\omega^2=\mathbf{|k|}^2$ and the photon group velocity remains $v_g=1$, while the amplitude changes. Thus, the prescribed scalar background does not change the geometric optics dispersion relation to next-to-leading order. However, there could be a modification of the real local photon  wavenumber due to the finite wavelength phase correction at higher orders.


\section{Scalar field radiation from an isolated compact star}\label{sec4}

In the situations considered in previous sections, the scalar field was coupled to a static source, resulting in a long-range $1/r$ scalar field profile outside the magnetized star at the far region. Since a static source correspond to the aligned rotator model does not emit any EM radiation, no scalar field radiation originates from the star in these situations.
To investigate the impact of scalar radiation on pulsar spin-down, we now consider a scenario where the EM fields of the magnetized star are oscillating with time. We consider a skewed rotator model, where the magnetic moment axis of the star makes an angle $\alpha$ with its rotation axis.
 
The leading quasi-static magnetic field can be written as \cite{deutsch}
\begin{equation}
\begin{aligned}
B_r &= \frac{ B_0 R^3}{r^3}\left(\cos\alpha \cos\theta+\sin\alpha\sin\theta\cos(\phi-\Omega t)\right),\\
B_\theta &= \frac{B_0 R^3}{2r^3}\left(\cos\alpha\sin\theta-\sin\alpha\cos\theta\cos(\phi-\Omega t)\right),\\
B_\phi &= \frac{B_0 R^3}{2r^3}\sin\alpha\sin(\phi-\Omega t),
\end{aligned}
\label{rad1}
\end{equation}
where $\alpha$ is the angle between the rotational axis and the magnetic moment axis, and $\theta$ is the polar angle with respect to the rotation axis. In the limit $\alpha\rightarrow 0$, Eq.~\eqref{rad1} reduces to Eq.~\eqref{eq:1m}. Thus, for radiation, one needs a non-zero $\alpha$.

The time-dependent source charge density for $\phi$ may be
written as $\rho_\phi(\mathbf{r},t)=g_{\phi\gamma\gamma}
(\mathbf{B}^2-\mathbf{E}^2)/2$, where Eq.~\eqref{rad1} gives
\begin{equation}
\mathbf{B}^2 - \mathbf{E}^2 \approx
\frac{3B_0^2 R^6}{2r^6}
\Big[\sin\alpha\cos\alpha\sin\theta\cos\theta
\cos(\Omega t - \phi)
+\frac{1}{2}\sin^2\alpha\sin^2\theta\cos^2(\Omega t - \phi)\Big],
\label{rad3}
\end{equation}
and we neglect terms suppressed by powers of $\Omega R$, and hence, $\mathbf{E^2}$. Subsequently, we also remove the time-independent terms which would not contribute to the radiation. 

In Eq.~\eqref{rad3}, the term $\mathbf{B}^2 - \mathbf{E}^2 \propto \sin\theta\cos\theta$ exhibits a quadrupolar angular dependence, while the $\sin^2\theta$ component contains both a time-independent DC part, which has been neglected later and a quadrupolar contribution. Consequently, the first term leads to quadrupolar radiation at the rotation frequency $\Omega$, whereas the second term produces radiation at twice the frequency, $2\Omega$. Therefore, unlike the pseudoscalar axion case \cite{Khelashvili:2024sup,Chadha-Day:2026ban}, no scalar dipole radiation arises in this scenario.

Therefore, the scalar-induced source charge density becomes
\begin{equation}
\rho_\phi=\frac{3g_{\phi\gamma\gamma}\mu^2}{r^6}\Big[\sin\alpha\cos\alpha\sin\theta\cos\theta
\cos(\Omega t - \phi)
+\frac{1}{2}\sin^2\alpha\sin^2\theta\cos^2(\Omega t - \phi)\Big],
\end{equation}
where $\mu=B_0R^3/2$. We adopt the quasi-static approximation for a slowly rotating compact star $(\Omega R\ll 1)$, which indicates that most of the quadrupole moment comes from the region where the near-zone approximation is appropriate. Since the scalar source density falls as $r^{-6}$, the dominant contribution to scalar radiation arises from near the stellar surface, where this approximation is valid. We therefore neglect corrections from the fully time-dependent EM field at large distances.

Using the convention $\rho(t)=\mathrm{Re}[\rho_\omega e^{i\omega t}]$, the radiating source amplitude at angular frequencies
$\Omega$ and $2\Omega$ can be written as
\begin{equation}
\rho_{+\Omega}(\mathbf{r})=\frac{3}{2}g_{\phi\gamma\gamma}\mu^2\sin(2\alpha)\sqrt{\frac{8\pi}{15}}\frac{Y_{2,-1}(\theta,\phi)}{r^6},~~~\rho_{+2\Omega}(\mathbf{r})=\frac{3}{4}g_{\phi\gamma\gamma}\mu^2\sin^2\alpha\sqrt{\frac{32\pi}{15}}\frac{Y_{2,-2}(\theta,\phi)}{r^6},
\end{equation}
where $Y_{l,m}(\theta,\phi)$  are spherical harmonics. 

The scalar field satisfies the wave equation
\begin{equation}
(\nabla^2+k^2)\phi_\omega(\mathbf{r})=\rho_\omega(\mathbf{r}),~~~k=\sqrt{\omega^2-m^2_\phi} \ ,
\end{equation} 
whose solution, using the retarded Green's function, is
\begin{equation}
\phi_\omega(\mathbf{r})=-\int d^3r^\prime \frac{e^{-ik |\mathbf{r}-\mathbf{r^\prime}|}}{4\pi|\mathbf{r}-\mathbf{r^\prime}|}\rho_\omega(\mathbf{r^\prime}).
\end{equation}  
Defining the multipole moments as
\begin{equation}
Q_{lm}(\omega)=\int d^3 r^\prime \rho_\omega(\mathbf{r}^\prime){r^\prime}^l Y_{l,m}^*(\hat{r}^\prime),
\end{equation}
the scalar field in the far field and long wavelength limit becomes
\begin{equation}
\phi_\omega(\mathbf{r})\simeq -\frac{e^{-ikr}}{r}\sum _{lm}i^l \frac{k^l}{(2l+1)!!}Y_{l,m}(\hat{r})Q_{lm}(\omega).
\end{equation}
For a real scalar field, the radial energy flux is
\begin{equation}
T^{0r}=-\dot{\phi}\partial_r\phi, ~~~\langle S_r\rangle=\langle T^{0r}\rangle=\frac{\omega k}{2}|\phi_\omega(\mathbf{r})|^2,
\end{equation}
where the angle brackets denote time-averaging over one period. Thus, the total power radiated in scalar waves can be obtained as
\begin{equation}
P_\omega=\int d\Omega r^2\langle S_r\rangle=\frac{\omega k}{2}\sum_{lm} \Big(\frac{k^l}{(2l+1)!!}\Big)^2|Q_{lm}(\omega)|^2.
\end{equation}
Hence, the scalar quadrupole radiation power at frequencies $\Omega$ and $2\Omega$ are given by
\begin{equation}
\begin{split}
P_\Omega\simeq\frac{\pi}{6000}g^2_{\phi\gamma\gamma}B^4_0R^{10}\Omega^6\Big(1-\frac{m^2_\phi}{\Omega^2}\Big)^{5/2}\sin^2 2\alpha\ \Theta(\Omega-m_\phi),\\ P_{2\Omega}\simeq\frac{4\pi}{375}g^2_{\phi\gamma\gamma}B^4_0 R^{10}\Omega^6\Big(1-\frac{m^2_\phi}{4\Omega^2}\Big)^{5/2}\sin^4\alpha \ \Theta(2\Omega-m_\phi),
\end{split}
\label{quadrad}
\end{equation}
where $\Theta$ denotes the Heaviside step function, which enforces the kinematic condition for propagating scalar radiation. Equation~\eqref{quadrad} denotes the leading quadrupole radiation result obtained within the near-zone, long-wavelength approximation. For $m_\phi < \Omega$, both $P_\Omega$ and $P_{2\Omega}$ contribute to scalar radiation and hence to the spindown luminosity. In the intermediate regime $\Omega < m_\phi < 2\Omega$, only the $P_{2\Omega}$ mode remains kinematically allowed and contributes to the energy loss. Scalar radiation arises only for a misaligned rotator, i.e., when the magnetic and rotation axes are not aligned; it vanishes in the aligned limit $\alpha \to 0$. For small inclination angles ($\alpha \lesssim 10^\circ$) and in the limit $m_\phi \ll \Omega$, the $P_\Omega$ mode dominates over $P_{2\Omega}$. 

\section{Constraints from Observations}\label{sec5}

\begin{figure}
\includegraphics[width=14cm]{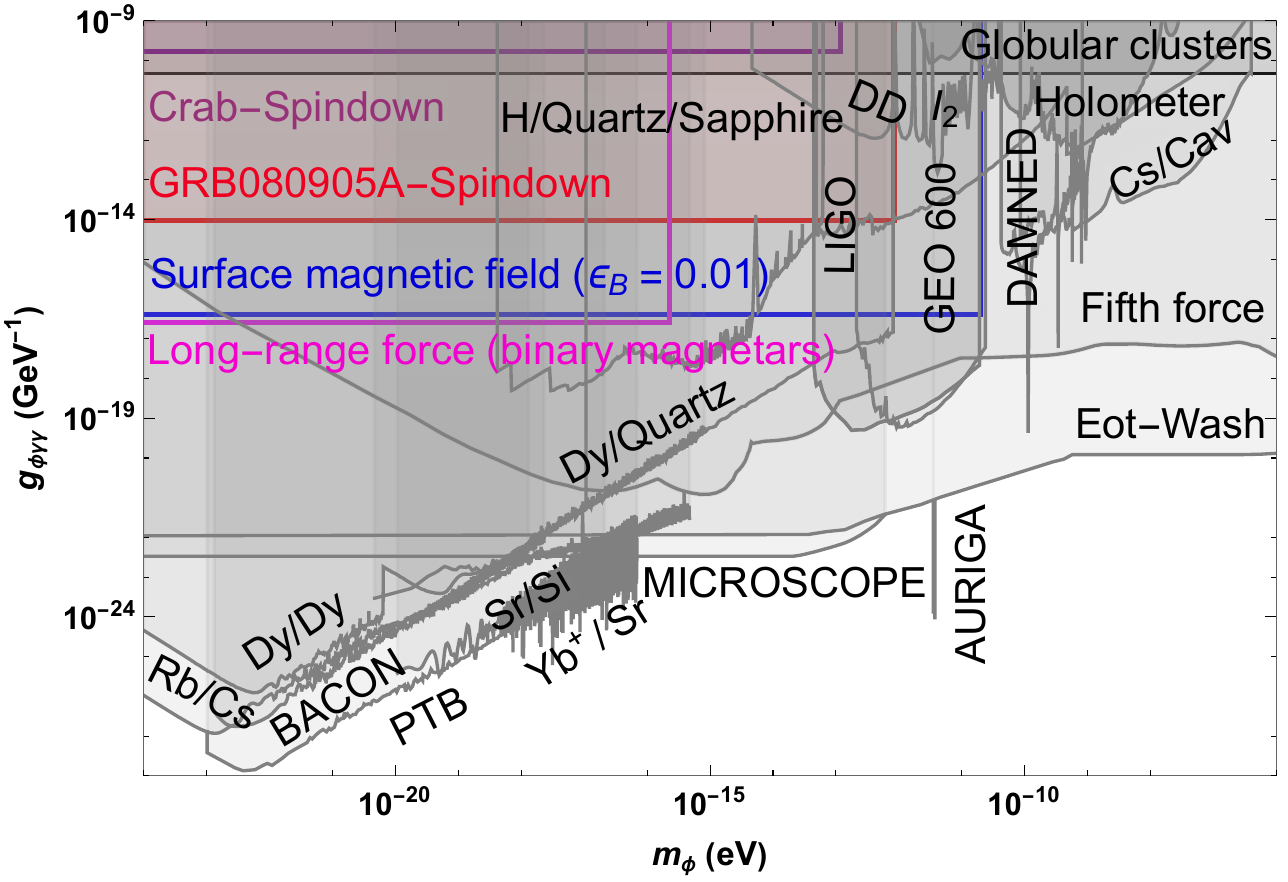}
\caption{Bounds and projected sensitivities on the scalar-photon coupling. The spindown constraints use data from the Crab pulsar and a GRB. The surface magnetic field constraints assume that the scalar contribution to the surface magnetic field is less that 1\% of its measured value. The (projected) constraints from the long-range force between binary magnetars have been obtained by assuming $B_{01}, B_{02} \sim 10^{17}$ G and the masses of the two magnetars known from independent measurements to within 0.05\% each.}
\label{plot8}
\end{figure}

In this section, we employ the results obtained in the previous sections and attain constraints and sensitivities on the scalar-photon coupling based on various observations related to pulsars and magnetars. 
The limits may originate from the sensitivities on the magnitude of a new long range force in double pulsar binaries, 
as well as from the measurements of the magnetic field of a compact star and of its spin-down luminosity.

A summary of the constraints on the scalar-photon coupling is given in \cite{AxionLimits}. Stringent bounds are obtained on the scalar-photon coupling by considering the scalar to be the DM. However, there are bounds that do not need the scalar to be the DM. The astrophysical bounds from globular clusters \cite{Dolan:2022kul} are obtained by examining the effects of additional scalar emission on stellar evolution, particularly through the observed number ratio of asymptotic giant branch to horizontal branch stars. The bounds from the E\"ot-Wash experiment \cite{Hees:2018fpg}, fifth force experiments~\cite{Fischbach:1996eq,Adelberger:2003zx,Konopliv2011,Fienga:2023ocw} and MICROSCOPE experiment \cite{Berge:2017ovy} are obtained from the precision tests of Einstein's equivalence principle, using laboratory measurements or observations in solar system. These bounds are relevant for comparison and complementarity with our work.

In Fig.~\ref{plot8}, the colored shaded regions represent the constraints on the scalar-photon coupling derived from pulsar spin-down luminosity measurements, while the colored curves indicate the projected sensitivities obtained from the prediction of the long-range force in binary systems and of the surface magnetic field of a compact star. For comparison, we also show the existing astrophysical constraints from globular clusters with gray shaded area bounded by the black line and other existing constraints from different observations represented by the gray shaded regions \cite{AxionLimits}.

\subsection{Search for a new long range force in a double pulsar binary}

We have seen in Section~\ref{sec1} that the scalar field interaction with the EM fields of a compact star induces a scalar charge at a far field region outside the star.
For a system of two compact stars in a binary, this would lead to a scalar-mediated long range force that has the same spatial dependence, $1/r^2$, as the gravitational force between the two stars.
The ratio of the long range force to the gravitational force is
\begin{equation}
\eta=\frac{4\pi Q^\mathrm{eff}_1 Q^\mathrm{eff}_2}{G M_1 M_2}\simeq \frac{4\pi g^2_{\phi\gamma\gamma}B^2_{01}B^2_{02}R^{3}_1 R^{3}_2}{144 GM_1M_2} \; ,
\label{eq:za1}
\end{equation}
where we use the expression for the scalar charge as given in Eq.~\eqref{eq:nm1} and write the Newton's gravitation constant  $G$ explicitly. The $4\pi$ factor arises from the normalization of the scalar Green's function and the relation between the field coefficient $Q_i^{\rm eff}$ and the integrated scalar source. Here $B_{01}$, $B_{02}$ are the surface magnetic fields of the two stars in a binary, $M_1$, $M_2$ are their masses, and $R_1$ and $R_2$ are their radii respectively.

As a concrete example to demonstrate how the scalar-mediated force may be constrained, we consider the double-pulsar binary system PSR J0737-3039 \cite{Kramer:2006nb,Kramer2008}. The polar magnetic fields are $B_{01}=2B_{1,\rm{eq}}$ and $B_{02}=2B_{2,\rm{eq}}$, where the surface magnetic fields of the two pulsars at the equator are $B_{1,\rm{eq}}\sim 6.3\times 10^{9}~\mathrm{G}$ and $B_{2,\rm{eq}}\sim 1.2\times 10^{12}~\mathrm{G}$ \cite{Lorimer:2007xw}. Their masses are $M_1=1.3381\pm 0.0007~M_\odot$ and $M_2= 1.2489\pm 0.0007~M_\odot$ \cite{Kramer:2006nb}. 
We take $R_1 = R_2 = 10$ km for an estimation.
This gives $\eta \sim (1.35\times 10^7) \ g^2_{\phi\gamma\gamma}~\mathrm{GeV}^2$.
The measurement/bound on $\eta$ can then be translated to the measurement/bound on $g_{\phi\gamma\gamma}$.

However, since the gravitational and the scalar-induced long range force have the same spatial dependence to the leading order at far region, it would be difficult to separate their contributions by simply measuring the attractive force between them. Indeed, a change in the magnitude of the force could be approximately mimicked by a change in the measured values of the masses. This quasi-no-go situation may be circumvented if the values of the masses are determined independently by some means other than the gravitational measurements or additional observables sensitive to differences in scalar charge-to-mass ratios  or we have access to a third body that is gravitationally bound with the binary but has a magnetic field much different than the two compact stars.

The distance between the two stars in PSR J0737-3039 is $a =8.8\times 10^5~\mathrm{km}$, which is known to $\sim 0.05 \%$. If indeed the masses of the stars in PSR J0737-3039 were also known to a precision of $\sim 0.05\%$ as the current measurements suggest, the precision in the prediction of the gravitational force would be $\sim 0.1\%$ and hence
the observations would be sensitive to $\eta \sim 10^{-3}$. The sensitivity on the coupling at this level results $g_{\phi\gamma\gamma} \sim 8.6\times 10^{-6}~\mathrm{GeV}^{-1}$.
Note that this value would be applicable only when the range of the scalar-induced force is more than the distance between the two compact stars, i.e. $1/m_\phi \gtrsim a$ or $m_\phi \lesssim 1/a\sim 2.2\times 10^{-16}~\mathrm{eV}$.

While our ability to obtain a concrete bound on
$g_{\phi\gamma\gamma}$ at this stage is limited by the lack of available information about the masses of the stars through non-gravitational means or from a third gravitationally-bound body, future observations may locate a system where these conditions are fulfilled.  Since the scalar charge is proportional to the square of the magnetic field strength, larger values of $B_0$ will give better constraints on $g_{\phi\gamma\gamma}$.
The constraint can be significantly improved for binary magnetar systems because of the larger magnetic fields. 
So far, no magnetar binary system has been detected. However, future experiments and observations with better sensitivity can explore this possibility \cite{Popov:2020ken,Chrimes:2022imi,Sherman:2024asw,Chandrasekhar:1953zz,Sinha:2010fm}.

For a benchmark, consider a hypothetical binary system consisting of two magnetars, each with a polar surface magnetic field of $B_0\sim 10^{16}~\mathrm{G}$, separated by the same distance as the components of PSR J0737-3039, and having identical masses and radii to the stars in that system. Let us also assume that the masses of the compact stars are measured with a precision of $0.05\%$, so that the measurements are sensitive to $\eta \sim 10^{-3}$. 
Under these conditions, the projected sensitivity on the scalar-photon coupling can be obtained as $g_{\phi\gamma\gamma}\sim 2.6\times 10^{-15}~\mathrm{GeV}^{-1}$.
If the polar surface magnetic fields were $10^{17}$ G each, the corresponding bound will be $g_{\phi\gamma\gamma}\sim 2.6\times 10^{-17}~\mathrm{GeV}^{-1}$. These prospective sensitivities have been given in Table~\ref{table2}. The last limit is represented by the magenta line in FIG. \eqref{plot8}. These values should be regarded as illustrative projections demonstrating the potential enhancement associated with very strongly magnetized compact stars, rather than precision bounds. A quantitative treatment in this regime would require incorporating strong field QED effects self consistently into the EM background.

\begin{table}[h]
\centering
\begin{tabular}{ |l|c|c|c|c|c|c| }
 \hline
 \multicolumn{4}{|c|}{Search for a new long-range force} \\
 \hline
Limits \hspace{0.01cm} & PSR J0737-3039\hspace{0.01cm}&$B_{01,02}\sim 10^{16}~\mathrm{G}$\hspace{0.01cm}&$B_{01,02}\sim 10^{17}~\mathrm{G}$\hspace{0.01cm}\\
 \hline
$g_{\phi\gamma\gamma}$ &$\sim 8.6\times 10^{-6}~\mathrm{GeV}^{-1}$ & $\sim 2.6\times 10^{-15}~\mathrm{GeV}^{-1}$&$\sim 2.6\times 10^{-17}~\mathrm{GeV}^{-1}$\\
\hline
\end{tabular}
\caption{\label{table2} Summary of the prospective sensitivities on the scalar-photon coupling. For compact stars separated by $a=8.8\times 10^5~\mathrm{km}$, these limits are valid for  $m_\phi \lesssim 1/a = 2.2 \times 10^{-16}$ eV.}
\end{table}
The effective scalar charge used in this analysis is obtained by solving the scalar field equation in a prescribed exterior EM background, without including the backreaction of the scalar field on the EM configuration. Consequently, the projected sensitivities are based on the leading order scalar charge within this fixed background approximation. For sufficiently large scalar-photon couplings, a self consistent determination of the scalar charge would require solving the coupled scalar and Maxwell equations.

\subsection{Searching for a scalar-induced magnetic field}
\label{subsecmag}

As discussed in Section~\ref{sec2}, the scalar field interaction with the EM fields of a compact star gives rise to a scalar-induced long range magnetic field $\mathbf{B}_\phi$. The observed surface magnetic field of the compact star at the pole has contributions from the standard EM fields and the scalar-induced magnetic field $\mathbf{B}_\phi|_{r=R}$ as given in Eq.~\eqref{final-scalar-mag}. Taking this into account,
the corresponding bound on $g_{\phi\gamma\gamma}$ can be obtained.

We obtain the magnitude of the scalar-induced magnetic field at the polar surface from Eq.~\eqref{final-scalar-mag} as
\begin{equation}
|B_\phi|=\frac{g^2_{\phi\gamma\gamma}B^3_0R^2}{120},     \label{measureb}
\end{equation}
whereas the perturbativity condition is given as
\begin{equation}
\epsilon_\phi=|g_{\phi\gamma\gamma}\phi(R,0)|=\frac{17}{240}g^2_{\phi\gamma\gamma}B^2_0R^2\ll1,
\label{perturbativity}
\end{equation}
following Eq.~\eqref{scalarsurfacemain1}. We parameterize the maximum allowed scalar-induced magnetic contribution as $\Delta B=\epsilon_B B_0$. Requiring $|B_\phi(R,0)|\leq \Delta B$ yields $g_{\phi\gamma\gamma}\lesssim \sqrt{120\epsilon _B}/B_0R$. Additionally, the perturbativity condition results $|g_{\phi\gamma\gamma}\phi|=8.5\epsilon_B$. Note that the quoted magnetic field intervals $B_0$ in Table~\ref{table4} do not universally represent statistical uncertainties and some reflect variations in stellar or magnetospheric modeling. Thus, to remain in the perturbative regime, we conservatively choose $\epsilon_B=0.01$ and obtain sensitivities on the scalar-photon coupling.

We use three sources for our analysis: the Crab pulsar \cite{Lyne1993,Bejger:2002ty,Manchester:2004bp,Philippov:2013aha}, the Soft Gamma Repeater SGR 1806-20 \cite{sp2010,Younes:2015hsa}, and GRB 080905A \cite{Rowlinson:2010jb,Rowlinson:2013ue}.
In Table~\ref{table4}, we summarize the input parameters of these compact stars such as their surface magnetic field at the pole $B_0$, and the radius of the compact star $R$, and also show the conservative sensitivities that we obtain on the scalar-photon coupling.
Note that these limits are valid when the range of the scalar field is more than the radius of the star, i.e. $1/m_\phi \gtrsim R$ or $m_\phi \lesssim 1/R$.

\begin{table}[h]
\centering
\begin{tabular}{ |c|c|c|c|c|c|c| }
 \hline
 \multicolumn{4}{|c|}{Search for a scalar-induced magnetic field} \\
 \hline
 \hspace{0.01cm} & Crab pulsar\hspace{0.01cm}&SGR 1806-20\hspace{0.01cm}&GRB 080905A\hspace{0.01cm}\\
 \hline
$B_0$ & $(3.4-4.0)\times 10^{12}~\mathrm{G}$ \cite{Khelashvili:2024sup} & $(1.6-5)\times 10^{15}~\mathrm{G}$ \cite{MEREGHETTI20111317} & $  (27.1-49.5)\times 10^{15}~\mathrm{G}$ \cite{Rowlinson:2013ue}\\
$R$ & $14~\mathrm{km}$ \cite{Khelashvili:2024sup} & $10~\mathrm{km}$ \cite{Marsden:2001tw}  &$10~\mathrm{km}$ \cite{Rowlinson:2013ue}\\
 \hline
$g_{\phi\gamma\gamma}$ &$\sim 2.3\times 10^{-13}~\mathrm{GeV}^{-1}$ & $\sim  6.9\times 10^{-16}~\mathrm{GeV}^{-1}$&$\sim 4.1 \times 10^{-17}~\mathrm{GeV}^{-1}$\\
$m_\phi$ & $\lesssim 1.4\times 10^{-11}~\mathrm{eV}$ &$\lesssim 2 \times 10^{-11}~\mathrm{eV}$ & $\lesssim 2 \times 10^{-11}~\mathrm{eV}$\\
\hline
\end{tabular}
\caption{\label{table4} Input Parameters for the candidate compact stars and sensitivities obtained on the scalar-photon coupling $g_{\phi\gamma\gamma}$, from the search for a scalar-induced magnetic field. These sensitivities are valid for the ranges of $m_\phi$ as shown.}
\end{table}

Observationally, the scalar-induced magnetic field may be probed through deviations from the radial and angular dependence of the conventional dipolar magnetic field. In principle, a joint analysis of magnetic field sensitive observables originating from different regions of the pulsar magnetosphere could help distinguish such an additional contribution. However, extracting this contribution would require a detailed modeling of the emission regions and the underlying magnetospheric magnetic field configuration.

We obtain the sensitivity on the scalar-photon coupling to be $g_{\phi\gamma\gamma}\sim 4.1\times 10^{-17}~\mathrm{GeV}^{-1}$  (valid for $m_\phi \lesssim 2 \times 10^{-11}$ eV), from the polar cap emission of GRB 080905A. This sensitivity is six orders of magnitude stronger than the astrophysical bound from globular clusters, which is $g_{\phi\gamma\gamma}\lesssim 4.7\times 10^{-11}~\mathrm{GeV}^{-1}$ \cite{Dolan:2022kul}.
This GRB sensitivity is depicted in Fig. \ref{plot8} by the blue curve. Note that, the sensitivity for Crab pulsar is two orders of magnitude stronger than the globular cluster bound and the sensitivity for SGR 1806-20 is also five order of magnitude stronger than the globular cluster constraint. However, since these two sensitivities are weaker than that from the GRB, we do not show them in the sensitivity plot.

\subsection{Spin-down of compact stars due to scalar radiation} 

The rotational energy of a compact star, and thus its spin, decreases due to the EM radiation. If these fields are time dependent, they result in radiated power, which decreases the rotational energy $E = I \Omega^2/2$ of the star, where $I$ is its moment of inertia.
The loss of rotational energy results in a decrease in $\Omega$ and hence an increase in the spin period $P$ of
the star,
\begin{equation}
L_{\mathrm{sd}} =\frac{dE}{dt} = \frac{d}{dt} \left(\frac{I}{2} \Omega^2 \right) \approx I \, \Omega \, \dot{\Omega} 
   = -I \frac{(2 \pi)^2}{P^3} \dot{P} \; .
\label{eq:Erot}
\end{equation}
We conservatively assume that the energy loss is entirely due to the magnetic dipole (MD) radiation, which is given as
\begin{equation}
L_{\mathrm{MD}}=-{\frac{dE}{dt}}\Bigg|_{\mathrm{MD}}
    = \frac{1}{6} (B_0 R^3\sin\alpha)^2 \Omega^4,
\label{eq:dipole}
\end{equation}
where $B_0$ is the polar surface magnetic field. From Eqs. \eqref{eq:Erot} and \eqref{eq:dipole},
we get
\begin{equation}
B_0\sin\alpha=\Big(\frac{3I}{2\pi^2 R^6}\Big)^{\frac{1}{2}}(P\dot{P})^{\frac{1}{2}} =6.4\times 10^{19}\sqrt{P\dot{P}}~\mathrm{G}\;,
\label{dew1}
\end{equation}
with a canonical choice of $I=10^{45}~\mathrm{g~cm^2}$, $R=10~\mathrm{km}$ and $P$ and $\dot{P}$ are measured in $\mathrm{s}$ and $\mathrm{s}~\mathrm{s^{-1}}$, respectively.

The GW radiation would also contribute, but its contribution would be negligible. The scalar radiation may also contribute to this spin-down, which can be measured using 
 $\dot{P}$, i.e., the rate of increase of the spin period $P$. The spin-down luminosity of the star is the rate of loss of its rotational energy, $dE/dt$, as given in Eq.~\eqref{eq:Erot}.
 
To constrain the scalar-photon coupling, we assume that the total energy loss rate of the pulsar receives contributions from both standard mechanisms and the scalar radiation given by Eq.~\eqref{quadrad}, where the scalar emission happens at frequencies $\Omega$ and $2\Omega$. It is worth noting that this mass range is narrower than the condition $m_\phi \lesssim 1/R$ discussed in the previous subsection, as for Crab pulsar $\Omega \sim 10^{-13}~\mathrm{eV}$ while $1/R \sim  10^{-11}~\mathrm{eV}$. The constraints on the scalar-photon coupling may be obtained by requiring $P_\phi\leq L^{\mathrm{obs}}_\mathrm{\mathrm{sd}}$, where the spin-down luminosity $L^{\mathrm{obs}}_\mathrm{\mathrm{sd}}$ for different stars is given in TABLE~\eqref{table3}. 

In practice, the scalar radiation from the magnetic dipole radiation can be indirectly distinguished by comparing the observed spin-down with the standard dipole prediction. The scalar radiation contribution to the energy loss scales as $\Omega^6$, compared with the $\Omega^4$ dependence of the conventional magnetic dipole emission. The scalar channel is characterized by a different dependence on $B_0$, $R$ and $\alpha$, together with the threshold factor $(1-m^2_\phi/\Omega^2)^{5/2}$, and can therefore be constrained through timing measurements, especially via deviations in the spin-down scaling or braking index.

In TABLE~\eqref{table3}, we summarize the bounds on scalar-photon coupling and scalar mass from the spin-down. The strongest bound on the coupling is obtained from the spin-down of GRB 080905A.


\begin{table}[h]
\centering
\begin{tabular}{ |c|c|c|c|c|c|c| }
 \hline
 \multicolumn{4}{|c|}{Spin-down of compact stars} \\
 \hline
 \hspace{0.01cm} & Crab pulsar\hspace{0.01cm}&SGR 1806-20\hspace{0.01cm}&GRB 080905A\hspace{0.01cm}\\
 \hline
$P$ &$33.8~\mathrm{ms}$ \cite{Lyne:2014qqa}&$7.468~\mathrm{s}$ \cite{Marsden:2001tw}& $9.80~\mathrm{ms}$ \cite{Rowlinson:2013ue}\\
 $\dot{P}$ &$4.2\times 10^{-13}~\mathrm{s~s^{-1}}$ \cite{Lyne:2014qqa}&$115.7\times 10^{-12}~\mathrm{s~s^{-1}}$ \cite{Marsden:2001tw}& $3.8\times 10^{-5}~\mathrm{s~s^{-1}}$\cite{Rowlinson:2013ue}\\
 $\alpha$ & $70^\circ$ \cite{Kou:2015oma} & $30^\circ$ \cite{Jia:2008ag} & $90^\circ$ \cite{Rowlinson:2010jb} \\
 $L^{\mathrm{obs}}_{\mathrm{sd}}$ & $7.9\times 10^{38}~\mathrm{erg}/\mathrm{s}$ & $1.1\times 10^{34}~\mathrm{erg}/\mathrm{s}$ & $1.6\times 10^{48}~\mathrm{erg}/\mathrm{s}$\\
 \hline
$g_{\phi\gamma\gamma}$ &$\lesssim 1.7\times 10^{-10}~\mathrm{GeV}^{-1}$ & $\lesssim 1.4\times 10^{-10}~\mathrm{GeV}^{-1}$&$\lesssim 9.5\times 10^{-15}~\mathrm{GeV}^{-1}$\\
$m_\phi$ & $\lesssim 1.2\times 10^{-13}~\mathrm{eV}$ &$\lesssim 5.5\times 10^{-16}~\mathrm{eV}$ & $\lesssim 8.4\times 10^{-13}~\mathrm{eV}$\\
\hline
\end{tabular}
\caption{\label{table3} Bounds obtained on the scalar-photon coupling $g_{\phi\gamma\gamma}$ for a range of scalar masses, from the measurements of spin-down.}
\end{table}

However, in case of GRB 080905A, the central engine period, magnetic field, and inferred $\dot P$ come from a magnetar model fitted to its afterglow, not direct pulsar timing. The model assumes a particular spin-down prescription, geometry and conversion of rotational energy to observable radiation. Thus, $L_{\rm sd}^{\rm obs}$ denotes the model-inferred initial spin-down luminosity.

In FIG. \eqref{plot8}, the constraint obtained from the spin-down of Crab pulsar and GRB 080905A are shown in the purple and red-shaded regions, while the bound from the SGR 1806-20 is not depicted, as it is of same order of magnitude with Crab. The bounds are valid for the mass of the scalar smaller than the respective spin angular frequencies of the star.
The Crab pulsar constraint is one order of magnitude weaker than the globular cluster bound and the GRB bound is stronger than the globular cluster bounds by two orders of magnitude. These are still weaker than the bounds from E\"ot-Wash \cite{Hees:2018fpg} and MICROSCOPE \cite{Berge:2017ovy} experiments. More sensitive pulsar spin-down measurements could improve our results by several orders of magnitude.

\section{Conclusions and Discussions}\label{sec6}

Ultralight scalar particles can couple with the 
time independent electric and magnetic fields of a compact star, which would result in a long range scalar field around the star with a spatial dependence $ \phi \sim 1/r$ at the far field region. The field profile is also calculated at the stellar surface which has a particular angular dependence.
Several laboratory and astrophysical measurements, such as the tests of the equivalence principle from E\"ot-Wash experiment, the fifth force experiments, and measurements in atomic spectroscopy, yield stringent constraints on the EM couplings of these scalars. In this paper, we propose and analyze multiple ways of constraining these couplings using astrophysical observations of pulsars, magnetars and double pulsar binaries.

We work in the context of a theory with an effective  scalar-photon coupling $g_{\phi\gamma\gamma}$. Such a coupling would typically arise from a UV theory with a mass scale $\Lambda$. 
In our work, $g_{\phi \gamma \gamma}$, and hence this $\Lambda$, are effective parameters and we are agnostic about their origin. The bounds obtained on $g_{\phi \gamma \gamma}$ can be translated to the bounds on $\Lambda$ in the context of a UV-complete theory, but it is beyond the scope of our work. The mass of the ultralight scalar, $m_\phi$,  may also be related to $\Lambda$ in the context of a UV-complete theory.
However, in the phenomenological work of ours, it is simply a free mass parameter.

The far field solution of $(\sim 1/r)$ spatial dependence of the scalar field differs from the $(\sim \cos\theta/r^2)$ dependence of the pseudoscalar axions that may couple to the EM field. Due to this spatial dependence, the scalar scenario may be considered equivalent to having a scalar charge $Q_\phi$ outside the star in the far region, giving rise to a Coulomb-like, spherically symmetric,  long range potential. In the near-field region as well as on the surface of the star, though, the scalar field has a finite angular dependence. For small values of $\Omega R$ (e.g. $\Omega R < 0.5$), the scalar field profile does not change substantially. The scalar-induced magnetic field has different radial and angular dependence compared to the standard dipole magnetic field --- unlike the dipole field, the scalar-induced magnetic field value is larger on the equator than at the poles. In the far region, the magnitude of the scalar-induced magnetic field is spherically symmetric. We further investigate the modification of the photon dispersion relation induced by the scalar-photon interaction. We find that the dispersion relation remains null at both leading and next-to-leading orders in the geometric optics expansion, although the photon wave amplitude is modified at next to leading order.

In a binary pulsar system where both of the compact stars give rise to scalar fields, the stars experience a long range scalar-mediated force in addition to gravity, arising from the interaction between the scalar fields and EM fields sourced by the two stars. The scalar-mediated force may be mimicked by a change in the masses of the two stars. However, if independent information about the masses of the two stars and the distance between them is available -- for example, from their interaction with a third body gravitationally bound to them but without a large EM field -- then it would be possible to detect this force or to constrain
its value. Using the parameters of the pulsar binary system PSR J0737-3039, if the masses of the two stars and the distance between them is known to a precision of $0.05\%$, we find that a sensitivity of $g_{\phi\gamma\gamma} \sim 8.6 \times 10^{-6}~\mathrm{GeV}^{-1}$ on the scalar-photon coupling may be obtained. This limit would be valid for a scalar mass of $m_\phi \lesssim 2.2 \times 10^{-16}~\mathrm{eV}$ so that the range of the force is more than the distance between the two stars in this binary system. 
This sensitivity is weaker than that from the existing globular cluster bound by several orders of magnitude. However, the coupling is inversely proportional to the product of the surface magnetic fields of each star, and future measurements of a binary magnetar system with a high magnetic field $(B_0\sim 10^{17}~\mathrm{G})$ could improve it to
$g_{\phi\gamma\gamma}\sim 2.6\times  10^{-17}~\mathrm{GeV}^{-1}$ if the masses of the stars and the distance between them are known to $\sim 0.05\%$. 

The surface magnetic field of a compact star can be inferred from measurements of its spin period and period derivative, while independent magnetic field observables can in principle constrain additional non standard contributions. Assuming that the polar surface magnetic field can be determined independently at the $1\%$ level, we find that the scalar-induced magnetic field of GRB 080905A could probe scalar-photon couplings down to $g_{\phi\gamma\gamma}\sim 4.1\times 10^{-17}~\mathrm{GeV}^{-1}$. This sensitivity applies for $m_\phi\lesssim 2\times 10^{-11}~\mathrm{eV}$, for which the scalar Compton wavelength exceeds the characteristic stellar radius and the long range scalar field solution remains applicable. The result illustrates the strong potential of highly magnetized compact stars for probing very weak scalar-photon interactions through deviations from the standard surface magnetic field configuration. The corresponding limit is consistently within the perturbativity regime.

If the background EM fields are time dependent, the scalar field will also be time-dependent and hence will radiate. This scalar radiation will lead to an additional spin-down of the star.
The spin-down luminosity, or the rate of change of rotational energy of the star, is a measurable quantity and can be obtained from the measurements of the spin period and its derivative. The scalar spin-down luminosity increases with increasing surface magnetic field, radius, and spin frequency of the star. We analyze the data on the Crab pulsar, SGR 1806-20, and GRB 080905A, and obtain the strongest bound on the scalar-photon coupling from the measurement of the spin-down luminosity of GRB 080905A as $g_{\phi\gamma\gamma}\lesssim 9.5\times 10^{-15}~\mathrm{GeV}^{-1}$ for $m_\phi\lesssim 8.4\times 10^{-13}~\mathrm{eV}$. The corresponding bound for the Crab pulsar is obtained as $g_{\phi\gamma\gamma}\lesssim 1.7\times 10^{-10}~\mathrm{GeV}^{-1}$, valid for $m_\phi\lesssim 1.2\times 10^{-13}~\mathrm{eV}$. 

The constraints and sensitivities discussed here from various observations can be further improved with enhanced sensitivity of detection and by focusing on stars with larger surface magnetic fields, larger radii, and higher spin frequencies. The strongest constraint on the scalar-photon coupling comes from the measurements of the spin-down luminosity of GRB, which is more than three orders of magnitude stronger than the existing globular cluster bound whereas the Crab pulsar bound is one order of magnitude weaker than that for the globular cluster. The strongest sensitivity on the coupling is obtained from the determination of the scalar-induced magnetic field for GRB as well as determination of the long-range force for a binary magnetar system.

The existing bounds on the scalar-photon coupling from the atomic clock experiments assume the scalar to be the DM, whereas our results do not require the scalar to be the DM. 
The MICROSCOPE bound, obtained through the measurement of the variation of nuclear binding energy induced by the change in the fine structure constant, is stronger. However, our bounds are obtained from astrophysical observations, without involving detailed nuclear physics considerations, and are therefore complementary to this bound. 
In future, developments such as advanced telescopes, combined with the detection of low frequency photon signals by instruments like LOFAR \cite{Geyer:2017lbi} and SKA \cite{Huege:2016jvc}, could enable precise measurements of deviations in the magnetic field measurement and precision gravitational force. This would potentially yield limits that are competitive with those from tests of equivalence principle experiments \cite{Berge:2017ovy}.

Note that the constraints and sensitivities presented in $g_{\phi\gamma\gamma}$ are actually on its magnitude. The expressions for the scalar-induced magnetic field, the energy loss rate from pulsars due to scalar radiation, and the scalar-mediated fifth force between two pulsars -- all involve even powers of the coupling $g_{\phi\gamma\gamma}$. As a result, the constraints and sensitivities derived on $g_{\phi\gamma\gamma}$ are insensitive to the sign of the coupling.

In deriving our results, we neglect the effects of the background plasma and the associated plasma-induced photon mass. In realistic magnetospheric environments, the plasma contribution may compete with the scalar-induced effects, potentially leading to distinctive observational signatures. In regions with extremely strong magnetic fields, quantum electrodynamical effects may also become important and could open additional radiation channels. The scalar-induced magnetic field could also be time-dependent if the scalar field is sourced by a time-dependent EM field. We will investigate such scenario in detail in a separate work. Our results further depend on the choice of the NS equation of state; however, such variations are not expected to significantly alter the order of magnitude of the obtained constraints.

Ultralight scalar particles arise in many theoretical models and motivate diverse experimental searches, regardless of their contribution to DM. We have explored their observable implications within a minimal framework, characterized solely by the scalar mass $m_\phi$ and its coupling to photons $g_{\phi\gamma\gamma}$, without invoking hidden sector assumptions. We point out for the first time that such scalars generate a long range monopole field profile outside the magnetized stars at the far region, and show that, unlike axions, they do not induce birefringence to the leading geometric optics order. Future precision measurements of pulsars and magnetars could pave the way for a better understanding of the physics of ultralight scalars, in addition to the astrophysics of these compact objects.

\section*{Acknowledgements}

The authors would like to thank Shadab Alam, Sumanta Chakraborty, Girish Kulkarni, Jamie Mcdonald, Arunava Mukherjee and Nicholas Rodd for useful discussions. The authors would also like to thank the anonymous referee for their fruitful comments and suggestions. T. K. P. would also like to thank the Galileo Galilei Institute for Theoretical Physics, ICTP, and University of Hamburg for the hospitality, the INFN for partial support during the completion and finalization of this work, and COST Actions COSMIC WISPers CA21106 and BridgeQG CA23130, supported by COST (European Cooperation in Science and Technology).
The work of A.D. is supported by the Department of Atomic Energy, Government of India, under Project Identification Number RTI 4012 and the J. C. Bose Grant ANRF/JBG/2025/000265/PS from the Anusandhan National Research Foundation, Government of India. 
 
\appendix
\section{Long-range scalar field outside a compact star}\label{app}

Here, we follow \cite{Goldreich:1969sb,Shapiro:1983du} to calculate the electric and magnetic field profiles for a compact star when its spin axis aligns with its magnetic dipole axis, to calculate analytic expressions that are valid even for relatively fast rotating stars. The dipolar magnetic field outside of a compact star is given in Eq.~\eqref{eq:1m}. The magnetic field just inside the surface of the star is given as 
\begin{equation}
\mathbf{B}^\mathrm{in}_{(r=R)} =B_0\Big(\cos\theta\hat{r}+\frac{\sin\theta}{2}\hat{\theta}\Big).
\label{ck1}
\end{equation}
If $\bm{\mathcal{J}}$ denotes the current density, then Ohm's law reads $\bm{\mathcal{J}}=\sigma(\mathbf{E}^\mathrm{in}+\mathbf{v}\times \mathbf{B}^\mathrm{in})$, where $\sigma$ denotes the conductivity and $\mathbf{v}$ denotes the velocity of the charges inside the star. Assuming that the NS is a perfect conductor ($\bm{\mathcal{J}}/\sigma\rightarrow 0$), we can write the Ohm's law as $\mathbf{E}^\mathrm{in} +(\mathbf{\Omega}\times\mathbf{r})\times\mathbf{B}^\mathrm{in}=0$, where $\mathbf{v}=\mathbf{\Omega}\times \mathbf{r}$. Using Eq.~\eqref{ck1}, we obtain the electric field just inside the surface of the NS as
\begin{equation}
\mathbf{E}^\mathrm{in}_{(r=R)}=-B_0\Big[\cos\theta \, (\mathbf{\Omega}\times \mathbf{r})\times\hat{r}+\frac{\sin\theta}{2}(\mathbf{\Omega}\times \mathbf{r})\times\hat{\theta}\Big].
\label{ck2}
\end{equation}
Since $\mathbf{v}=\Omega R\sin\theta\, \hat{\phi}$, Eq.~\eqref{ck2} becomes
\begin{equation}
\mathbf{E}^\mathrm{in}_{(r=R)}=B_0\Omega R\sin\theta\Big(\frac{\sin\theta}{2}\hat{r}-\cos\theta\,\hat{\theta}\Big).
\label{ck3}
\end{equation}
As the tangential component of the electric field is continuous at $r=R$, from Eq.~\eqref{ck3} we obtain
\begin{equation}
\mathbf{E}^\mathrm{out}_{\theta (r=R)}=-\frac{\partial}{\partial\theta}\Big(\frac{B_0\Omega R\sin^2\theta}{2}\Big)=\frac{\partial}{\partial\theta}\Big(\frac{B_0\Omega R}{3}P_2(\cos\theta)\Big),
\label{ck4}
\end{equation}
where $P_2(\cos\theta)=\frac{1}{2}(3\cos^2\theta-1)$ is the Legendre polynomial of degree $2$. Assuming the outer region of the star is vacuum, we can write $\mathbf{E}^\mathrm{out}=-\nabla\Phi$, where  $\nabla^2\Phi=0$, correspond to the Laplace's equation. Using the boundary condition Eq.~\eqref{ck4} at $r=R$, the solution of Laplace's equation becomes
\begin{equation}
\Phi=-\frac{B_0\Omega R^5}{3r^3}P_2(\cos\theta),
\label{ck5}
\end{equation}
where we consider that the star has zero net electric charge. Thus, the scalar potential is quadrupolar in nature. Using Eq.~\eqref{ck5}, we obtain the expression for the electric field profile outside of the compact star as given in Eq.~\eqref{eq:2m}.  

To solve Eq.~\eqref{eq:5m} when $\Omega$ is not negligibly small, it is crucial to include the contribution from $\mathbf{E}^2$, since $\mathbf{E}^2$ in Eq.~\eqref{eq:3m} contains terms proportional to $(\Omega R)^2$.
Therefore, we use the Green's function method for solving the inhomogeneous differential equation Eq.~\eqref{eq:5m}. 
To obtain the static scalar field profile outside the star in flat space, we write the Poisson's equation 
\begin{equation}
\nabla^2\phi(r,\theta)=\rho_\phi(r,\theta),  
\label{dur1}
\end{equation}
where the scalar-induced source charge density is
\begin{equation}
\rho_\phi(r,\theta)=g_{\phi\gamma\gamma}\frac{B^2_0R^6}{8r^6}(3\cos^2\theta+1)-g_{\phi\gamma\gamma}\frac{B^2_0\Omega^2R^{10}}{8r^8}(5\cos^4\theta-2\cos^2\theta+1),~~~r>R,
\label{eq:8}
\end{equation} 

outside the star ($r \geq R$). In the following, we calculate the magnitude of the scalar field profile for the above scalar-induced source charge density. The boundary conditions are $\phi(r)$ must remain finite as $r\to 0$, the scalar field vanishes at spatial infinity, i.e., $\phi(r,\theta)\to 0$ as $r\to\infty$. Also, the scalar field and its radial derivative are continuous at the stellar surface. The scalar field and its radial derivative are also continuous at the stellar surface.

Using the exact Legendre decomposition, we can write Eq.~\eqref{eq:8} as
\begin{equation}
\rho_\phi(r,\theta)=g_{\phi\gamma\gamma}\frac{B^2_0R^6}{8r^6}\Big[2P_0(\mu)+2P_2(\mu)\Big]-g_{\phi\gamma\gamma}\frac{B^2_0 \Omega^2R^{10}}{8r^8}\Big[\frac{4}{3}P_0(\mu)+\frac{32}{21}P_2(\mu)+\frac{8}{7}P_4(\mu)\Big],
\label{dur16}
\end{equation}
where the $l=0,2,4$ terms contribute with the coefficients
\begin{equation}
{\rho_\phi}_0(r)=\mathcal{C}\Big(\frac{R^6}{2r^6}-\frac{\Omega^2R^{10}}{3r^8}\Big), ~~~{\rho_\phi}_2(r)=\mathcal{C}\Big(\frac{R^6}{2r^6}-\frac{8\Omega^2R^{10}}{21r^8}\Big),~~~{\rho_\phi}_4(r)=-\mathcal{C}\Big(\frac{2\Omega^2R^{10}}{7r^8}\Big),
\label{dur17}
\end{equation}
where $\mathcal{C}=g_{\phi\gamma\gamma}B^2_0/2$.

The static Green's function $G(\mathbf{x},\mathbf{x}^\prime)$ for the 3D Laplacian satisfies
\begin{equation}
\nabla^2G(\mathbf{x},\mathbf{x}^\prime)=\delta^{(3)}(\mathbf{x}-\mathbf{x}^\prime),~~~G(\mathbf{x},\mathbf{x}^\prime)=-\frac{1}{4\pi|\mathbf{x}-\mathbf{x}^\prime|},
\label{dur2}
\end{equation}
and one can obtain the solution of the scalar field as 
\begin{equation}
\phi(\mathbf{x})=\int d^3x^\prime G(\mathbf{x},\mathbf{x}^\prime)\rho_\phi(\mathbf{x}^\prime)=
-\frac{1}{4\pi}\int d^3x^\prime \frac{\rho_\phi(\mathbf{x}^\prime)}{|\mathbf{x}-\mathbf{x}^\prime|}.
\label{dur3}
\end{equation}

The scalar field solution Eq.~\eqref{dur3} can be expressed in Legendre polynomials $P_l(\cos\gamma)$ with degree $l$, considering $r=|\mathbf{x}|$, $r^\prime=|\mathbf{x^\prime}|$, and $\gamma$ is the angle between $\mathbf{x}$ and $\mathbf{x}^\prime$ so that
\begin{equation}
\frac{1}{|\mathbf{x}-\mathbf{x}^\prime|}=\sum_{l=0}^\infty \frac{r^l_<}{r^{l+1}_>}P_l (\cos\gamma), 
\label{dur4}
\end{equation}
and hence
\begin{equation}
\phi(\mathbf{x})=-\frac{1}{4\pi}\sum_{l=0}^\infty d^3x^\prime \rho_\phi(\mathbf{x^\prime}) \frac{r^l_<}{r^{l+1}_>}P_l (\cos\gamma),  
\label{dur5}
\end{equation}
where $r_<\equiv \mathrm{min}(r,r^\prime)$, $r_>\equiv \mathrm{max}(r,r^\prime)$, and $\cos\gamma=\cos\theta\cos\theta^\prime+\sin\theta\sin\theta^\prime\cos(\varphi-\varphi^\prime)$.

Since the scalar-induced source term $\rho_\phi(r,\theta)$ does not depend on $\varphi^\prime$, the integration over $\varphi^\prime$ projects out the axisymmetric part of the Green's function. Therefore, we can write
\begin{equation}
\int ^{2\pi}_0 d\varphi^\prime P_l(\cos\gamma)=2\pi P_l(\cos\theta)P_l(\cos\theta^\prime).
\label{dur6}
\end{equation}
Using Eq.~\eqref{dur6}, the scalar field profile given in Eq.~\eqref{dur5} becomes
\begin{equation}
\phi(r,\theta)= -\frac{1}{4\pi}\sum _{l=0}^\infty[2\pi P_l(\cos\theta)]\int ^\infty_0 dr^\prime {r^\prime}^2\frac{r^l_<}{r^{l+1}_>}\int ^\pi_0 d\theta^\prime\sin\theta^\prime P_l(\cos\theta^\prime)\rho_\phi(r^\prime,\theta^\prime),
\label{dur7}
\end{equation}
where we have defined the Legendre coefficients of the source 
\begin{equation}
{\rho_\phi}_l(r^\prime)\equiv \frac{2l+1}{2}\int ^{+1}_{-1}d\mu^\prime P_l(\mu^\prime)\rho_\phi(r^\prime,\mu^\prime),~~~\mu^\prime=\cos\theta^\prime.
\label{dur8}
\end{equation}
Therefore, Eq.~\eqref{dur7} simplifies to the standard mode expansion as
\begin{equation}
\phi(r,\theta)=\sum_{l=0}^\infty\phi_l(r)P_l(\cos\theta),   
\label{dur10}
\end{equation}
with 
\begin{equation}
\phi_l(r)=-\frac{1}{2l+1}\int ^\infty_R dr^\prime {r^\prime}^2 {\rho_\phi}_l(r^\prime)\frac{r^l_<}{r^{l+1}_>}. 
\label{dur11}
\end{equation}
Since the scalar-induced source charge density is only outside the star, i.e., ${\rho_\phi}_l(r^\prime)=0$ for $r^\prime<R$, therefore the lower limit for the radial integration starts with $r^\prime=R$.

Thus, the Green's function solution for $l$'th mode becomes
\begin{eqnarray}
\phi_l(r)&=& -\frac{1}{2l+1}\Big[\int ^r_R dr^\prime {r^\prime}^2 {\rho_\phi}_l(r^\prime)\frac{{r^\prime}^l}{r^{l+1}}+\int^\infty_r dr^\prime {r^\prime}^2 {\rho_\phi}_l(r^\prime)\frac{r^l}{{r^\prime}^{l+1}}\Big]\nonumber\\
&=& -\frac{1}{2l+1}\Big[\frac{1}{r^{l+1}}\int^r_R dr^\prime {r^\prime}^{l+2}{\rho_\phi}_l(r^\prime)+r^l\int^\infty_r dr^\prime {r^\prime}^{1-l} {\rho_\phi}_l(r^\prime) \Big].
\label{dur12}
\end{eqnarray}

The full exterior solution of the scalar field profile can be calculated using Eq.~\eqref{dur12}. Note that since $\rho_{\phi_l}$ is nonzero only for $l=0,2,4$, all other multipoles of $\phi_l$ vanish. We start with the $l=0$ contribution to Eq.~\eqref{dur12} and obtain
\begin{equation}
\phi_0(r)=-\mathcal{C}\Big[\frac{R^3}{6r}-\frac{R^6}{24r^4}-\frac{\Omega^2R^5}{15r}+\frac{\Omega^2 R^{10}}{90 r^6}\Big].    
\label{mono}
\end{equation}
The quadrupole contribution $(l=2)$ becomes
\begin{equation}
\phi_2(r)=-\mathcal{C}\Big[\frac{R^5}{10r^3}-\frac{R^6}{12r^4}-\frac{8\Omega^2R^7}{315 r^3}+\frac{\Omega^2 R^{10}}{63r^6}\Big].
\label{quad}
\end{equation}
Finally, the hexadecapole contribution $(l=4)$ leads to 
\begin{equation}
\phi_4(r)=-\mathcal{C}\Big[-\frac{2\Omega^2 R^9}{63 r^5}+\frac{\Omega^2 R^{10}}{35 r^6}\Big].
\label{hexa}
\end{equation}

Thus, the total exterior solution of the scalar field can be obtained using Eqs.~\eqref{mono}, \eqref{quad}, \eqref{hexa} and \eqref{dur10} as
\begin{align}
\begin{split}
\phi(r,\theta)&=-g_{\phi\gamma\gamma}B^2_0\Bigg\{\Big[\frac{R^3}{12r}-\frac{R^6}{48r^4} -\frac{\Omega^2 R^5}{30 r}+\frac{\Omega^2 R^{10}}{180 r^6}\Big] +\\
&\Big[\frac{R^5}{20r^3}-\frac{R^6}{24r^4}-\frac{4\Omega^2 R^7}{315r^3}+\frac{\Omega^2R^{10}}{126 r^6}\Big] P_2(\cos\theta)
+\Big[-\frac{\Omega^2R^9}{63r^5}+\frac{\Omega^2 R^{10}}{70 r^6}\Big]P_4(\cos\theta)\Bigg\}. 
\end{split}
\label{total-main}
\end{align}
In the far field region $r\gg R$, Eq.~\eqref{total-main} reduces to
\begin{equation}
\phi(r,\theta)\simeq -\frac{g_{\phi\gamma\gamma}B^2_0R^3}{12r}+\frac{g_{\phi\gamma\gamma}B^2_0\Omega^2R^5}{30r}, ~~~r\gg R. 
\label{durA20}
\end{equation}
To calculate the scalar field profile on the stellar surface, we take $r\to R$, which gives
\begin{equation}
\phi(R,\theta)=-g_{\phi\gamma\gamma}B^2_0R^2\Big[\frac{1}{16}+\frac{1}{120}P_2(\cos\theta)-(\Omega R)^2\Big(\frac{1}{36}+\frac{1}{210}P_2(\cos\theta)+\frac{1}{630}P_4(\cos\theta)\Big)\Big].   
\label{surfaceA4}
\end{equation}
For a slowly rotating star $\Omega R\ll1$, the surface field profile in Eq.~\eqref{surfaceA4} reduces to
\begin{equation}
\phi(R,\theta)\simeq -g_{\phi\gamma\gamma}B^2_0R^2\Big(\frac{7}{120}+\frac{\cos^2\theta}{80}\Big).   
\label{surfs}
\end{equation}
Thus, even for a slowly rotating star, the surface scalar field is not spherically symmetric. In the limit $r\to R$ and $\Omega R\ll 1$, Eq.~\eqref{total-main} reduces to Eq.~\eqref{surfs}.

\section{Scalar-induced magnetic field profile}\label{appnew2}
The scalar-induced magnetic field profile can be obtained by solving 
\begin{equation}
\nabla^2 \mathbf{B}_\phi= g_{\phi\gamma\gamma}\nabla\times (\nabla\phi\times \mathbf{B}_{(0)}).
\label{mb1}
\end{equation}

To calculate the scalar-induced magnetic field outside the star, we start with Eq.~\eqref{total-main} for a slowly rotating star $\Omega R\ll 1$ as
\begin{equation}
\phi(r,\theta)=-g_{\phi\gamma\gamma}B^2_0\Big[\frac{R^3}{12 r}-\frac{R^6}{48 r^4}+\Big(\frac{R^5}{20 r^3}-
\frac{R^6}{24r^4}\Big)P_2(\cos\theta)\Big].
\label{newapp1}
\end{equation}
Equation.~\eqref{newapp1} is necessary to calculate the scalar-induced magnetic field at the surface as $r^{-3}$ and $r^{-4}$ terms cannot be neglected near the surface.

We write the background dipole magnetic field as $\textbf {B}_{(0)} = B^r_{(0)} \hat{\bm{r}} + B^\theta_{(0)}\hat{\bm \theta} $, with
\begin{equation}
B^r_{(0)}=B_0R^3\frac{\cos\theta}{r^3}, ~~~B^\theta_{(0)}=\frac{B_0R^3}{2}\frac{\sin\theta}{r^3}.  
\label{newapp2}
\end{equation}
Since both $\nabla\phi$ and $\mathbf{B}_{(0)}$ lie in the $r-\theta$ plane,
\begin{equation}
\nabla\phi\times \mathbf{B}_{(0)}=Z_\varphi \hat{\bm{\varphi}},~~~Z_\varphi=\frac{\partial\phi}{\partial r}B^\theta_{(0)}-\frac{1}{r}\frac{\partial\phi}{\partial \theta}B^r_{(0)},  
\label{newapp3}
\end{equation}
which is necessary to calculate the right hand side of Eq.~\eqref{mb1}. Using Eqs.~\eqref{newapp1} and \eqref{newapp2}, we obtain
\begin{equation}
Z_\varphi=-\frac{g_{\phi\gamma\gamma}B^3_0R^6}{240 r^7}[9R^2(1+\cos^2\theta)-10r^2]\sin\theta.
\label{newapp4}
\end{equation}
Then Eq.~\eqref{mb1} gives
\begin{equation}
\nabla^2 \mathbf{B}_\phi=S_r\hat{\bm{r}}+S_\theta \hat{\bm{\theta}},  
\label{newapp7}
\end{equation}
where
\begin{equation}
\begin{split}
S_r=-\frac{g^2_{\phi\gamma\gamma}B^3_0R^6}{60 r^8}(9R^2\cos^2\theta-5r^2)\cos\theta,~~
S_\theta=-\frac{g^2_{\phi\gamma\gamma}B^3_0R^6}{120 r^8}[27R^2(1+\cos^2\theta)-20r^2]\sin\theta.
\end{split}
\label{newapp8}
\end{equation}
Defining $K=g^2_{\phi\gamma\gamma}B^3_0 R^6$ and using Legendre polynomial identities, Eq.~\eqref{newapp8} becomes
\begin{equation}
\begin{split}
S_r=-K\Big[\frac{27R^2-25r^2}{300 r^8}P_1(\cos\theta)+\frac{3R^2}{50r^8}P_3(\cos\theta)\Big],~~
S_\theta=-K\Big[\frac{50r^2-81R^2}{300r^8}\frac{dP_1}{d\theta}-\frac{3R^2}{100r^8}\frac{dP_3}{d\theta}\Big].
\end{split}
\label{newapp9}
\end{equation}

To solve the scalar-induced magnetic field, we use the multipole ansatz
\begin{equation}
\mathbf{B}_\phi=\sum_{l=1,3}\Big[b_l(r)P_l(\cos\theta)\hat{\bm{r}}+c_l(r)\frac{dP_l(\cos\theta)}{d\theta}\hat{\bm{\theta}}\Big].
\label{newapp10}
\end{equation}
Since scalar-induced magnetic fields are divergence-free, i.e., $\nabla\cdot \mathbf{B}_\phi=0$, we obtain from Eq.~\eqref{newapp10} 
\begin{equation}
c_l(r)=\frac{1}{l(l+1)r}\frac{d}{dr}[r^2b_l(r)].    
\label{newapp11}
\end{equation}
The radial component of the vector Laplacian can be obtained from Eq.~\eqref{newapp10} as
\begin{equation}
(\nabla^2 \mathbf{B}_\phi)_r=\Big[b_l^{\prime\prime}+\frac{2}{r}b_l^\prime-\frac{l(l+1)+2}{r^2}b_l+\frac{2l(l+1)}{r^2}c_l\Big]P_l.  \label{newapp12}
\end{equation}
Using Eq.~\eqref{newapp11}, the above equation simplifies to 
\begin{equation}
b^{\prime\prime}_l+\frac{4}{r}b^\prime_l+\frac{2-l(l+1)}{r^2}b_l=S_{r,l}.  
\label{newapp13}
\end{equation}
We start with the $l=1$ contribution and the corresponding contribution from Eq.~\eqref{newapp9} becomes
\begin{equation}
S_{r,1}=-K\Big(\frac{9R^2}{100r^8}-\frac{1}{12r^6}\Big).    
\label{newapp14}
\end{equation}
For $l=1$, Eq.~\eqref{newapp13} becomes
\begin{equation}
b_1^{\prime\prime}+\frac{4}{r}b_1^\prime=-K\Big(\frac{9R^2}{100r^8}-\frac{1}{12r^6}\Big).  
\label{newapp15}
\end{equation}
The decaying particular exterior solution $b_1(r)$ and hence $c_1(r)$ becomes
\begin{equation}
b_1(r)= \frac{K}{48 r^4}-\frac{KR^2}{200 r^6}, ~~~c_1(r)=-\frac{K}{48r^4}+\frac{KR^2}{100 r^6}
\label{newapp16}
\end{equation}
Similarly, the $l=3$ contribution of Eq.~\eqref{newapp9} is
\begin{equation}
S_{r,3}=-\frac{3KR^2}{50r^8}, \label{newapp18}   
\end{equation}
and Eq.~\eqref{newapp13} becomes
\begin{equation}
b_3^{\prime\prime}+\frac{4}{r}b^\prime_3-\frac{10}{r^2}b_3=-\frac{3KR^2}{50r^8}.
\label{newapp19}
\end{equation}
The decaying particular exterior solution $b_3(r)$ and consequently $c_3(r)$ becomes
\begin{equation}
b_3(r)=-\frac{3KR^2}{400 r^6},~~~c_3(r)=\frac{KR^2}{400r^6}.
\label{newapp20}
\end{equation}

Thus, the radial part of the scalar-induced magnetic field outside of the star becomes
\begin{equation}
B_\phi^r=\Big(\frac{K}{48r^4}-\frac{KR^2}{200r^6}\Big)P_1(\cos\theta)- \frac{3KR^2}{400r^6}P_3(\cos\theta). 
\label{newapp22}
\end{equation}
Similarly, the angular part of the scalar-induced magnetic field outside of the star becomes
\begin{equation}
B^\theta_\phi=\Big(-\frac{K}{48r^4}+\frac{KR^2}{100r^6}\Big)\frac{dP_1}{d\theta}+\frac{KR^2}{400r^6}\frac{dP_3}{d\theta}.
\label{newapp23}
\end{equation}
The total scalar-induced magnetic field particular solution exterior to the star is then
\begin{equation}
\begin{split}
\mathbf{B}_\phi(r,\theta)=g^2_{\phi\gamma\gamma}B^3_0R^6\Bigg\{\Big[\frac{1}{48r^4}-\frac{R^2}{160r^6}(3\cos^2\theta-1)\Big]\cos\theta\,\hat{\bm{r}}+ \\\Big[\frac{1}{48r^4}-\frac{R^2}{160r^6}(1+3\cos^2\theta)\Big]\sin\theta\,\hat{\bm{\theta}}\Bigg\}. 
\end{split}
\label{newapp24}
\end{equation}

In the far-field region $r\gg R$, Eq.~\eqref{newapp24} becomes
\begin{equation}
\mathbf{B}_\phi=\frac{g^2_{\phi\gamma\gamma}B^3_0R^6}{48r^4}(\cos\theta\,\hat{\bm{r}}+\sin\theta\,\hat{\bm{\theta}}),  
\label{newapp25}
\end{equation}
Similarly, on the stellar surface, the value of the scalar-induced magnetic field becomes
\begin{equation}
\mathbf{B}_\phi(R,\theta)= \frac{g^2_{\phi\gamma\gamma}B^3_0R^2}{480}[(13-9\cos^2\theta)\cos\theta\,\hat{\bm{r}}+(7-9\cos^2\theta)\sin\theta\,\hat{\bm{\theta}}]. 
\label{newapp26}
\end{equation}

Note that we have focused on the particular solution of Eq.~(\ref{mb1}) sourced by the scalar-photon interaction and set the homogeneous solutions to zero. This prescription excludes independent corrections to the asymptotic magnetic dipole and higher multipoles. Their determination would require additional boundary conditions and matching to the stellar interior.

\bibliographystyle{JHEP}
\bibliography{reference_scalar}
\end{document}